\documentclass[10pt, pre, amsmath, onecolumn, showpacs, superscriptaddress]{revtex4-1}
\usepackage{amsmath}
\usepackage{amssymb}
\usepackage[mathscr]{eucal}

\usepackage{graphicx}
\usepackage{latexsym}

\usepackage{color}
\definecolor{dgreen}{rgb}{0,0.7,0}

\newcommand{\ee}{\mathrm{e}}

\newcommand{\st}{\mathrm{st}}
\newcommand{\cZ}{\mathcal{Z}}

\newcommand{\hw}{\hat{w}}

\begin{document}
\title{Exact gap statistics for the random average process on a ring with a tracer} 
\author{J. Cividini$^1$, A. Kundu$^{1,2}$, Satya N. Majumdar$^3$ and D. Mukamel}

\affiliation{Department of Physics of Complex Systems, Weizmann Institute of Science, Rehovot 76100, Israel \\
$^2$International center for theoretical sciences, TIFR, Bangalore - 560012, India\\
$^3$LPTMS, CNRS, Univ. Paris-Sud, Universit\'e Paris-Saclay, 91405 Orsay, France.}

\begin{abstract}
We study statistics of the gaps in Random Average Process (RAP) on a ring with particles hopping symmetrically, 
except one tracer particle which could be driven. These particles hop either to the left or to the right by a random fraction $\eta$ of the 
space available till next particle in the respective directions. The random fraction $\eta \in [0,~1)$ is chosen from a distribution $R(\eta)$.
For non-driven tracer, when $R(\eta)$ satisfies a necessary and sufficient condition, the 
stationary joint distribution of the gaps between successive particles takes an universal form that is factorized except for a global constraint. 
Some interesting explicit forms of $R(\eta)$ are found which satisfy this condition.
In case of driven tracer, the system reaches a current-carrying steady state where such factorization does not hold. 
Analytical progress has been made in the thermodynamic limit, where we computed the single site mass distribution inside the bulk. 
We have also computed the two point gap-gap correlation exactly in that limit. Numerical simulations support our analytical results.
\end{abstract}

\maketitle

\section{Introduction}

Interacting particle problems have been widely studied in statistical mechanics as examples of many-body systems 
that usually exhibit rich types of behavior~\cite{spohn1991}.  
For equilibrium systems all the stationary properties are in principle known through the Gibbs-Boltzmann distribution, 
and the dynamics close to equilibrium can be described in very general terms~\cite{bertini2015}. 
Out of equilibrium, studying such interacting many-particle systems analytically in dimensions other than one is usually difficult.
In one dimension, one of the most studied example is the Simple Exclusion Process (SEP), where each lattice site is occupied 
by one hardcore particle or it is empty. In every small time interval $dt$, 
each particle moves to the neighboring site on the right (left) with probability $pdt$ ($qdt$) iff the target site is empty. 
A large number of results are known for this system (see \cite{derrida2007, chou_m_z2011} and references therein) and 
these are obtained using sophisticated analytical tools 
such as Bethe Ansatz, Matrix Product Ansatz or mappings to growth models~\cite{chou_m_z2011, lazarescu2015}. 
Another widely studied model of interacting particle systems is the \textit{Random Average Process} (RAP).
In the RAP particles move on a one dimensional continuous line in contrast to SEP where hardcore particles move on a lattice. Each particle moves to 
the right (left) by a \textit{random fraction} of the space available till the next nearest particle on the right (left) with some rate. 
Thus the jumps in one direction is a random fraction $\eta$ of the \textit{gap} to the nearest particle in that direction where the random number 
$\eta \in [0,1)$ is chosen from some distribution $R(\eta)$.

Many different results have been reported for this model in the literature. For example, dynamical properties like diffusion coefficient, 
variance of the tracer position and the 
pair correlation between the positions of two particles have been studied on an infinite line ~\cite{krug_g2000, schutz2000, rajesh_m2001}. 
In particular, it has been shown that the variance of the tagged
particle position in the steady state grows at late times as $\sim t$ for
the asymmetric case (jumping rates to the right or to the left are different), 
whereas for the symmetric case it grows as $\sim \sqrt{t}$. For the fully asymmetric case, this result was first derived by
Krug and Garcia \cite{krug_g2000} using a heuristic hydrodynamic description. Later, the
variance as well as other correlation functions, both for the symmetric
and asymmetric cases, were computed rigorously by Rajesh and Majumdar \cite{rajesh_m2001}.
The RAP model has also been shown to be linked to the porous medium equation~\cite{feng_i_s1996} and appears in a variety of problems 
like the force propagation in granular media~\cite{coppersmith1996, rajesh_m2000}, in models of 
mass transport \cite{rajesh_m2000, krug_g2000}, models of voting systems~\cite{melzak1976, ferrari_f1998}, models of wealth distribution~\cite{ispolatov1998}
and in the generalized Hammersley process~\cite{aldous_d1995}. The RAP can be shown to be equivalent, up to an overall translation, to Mass Transfer (MT)
 models~\cite{krug_g2000} when one identifies 
the gaps or the inter-particle spacings in the RAP picture with the masses. In terms of MT picture, statistics of masses/gaps has 
been well studied. For uniform distributions $R(\eta)$, invariant measures of the masses have been computed for a general 
class of RAP (or equivalently MT) models~\cite{rajesh_m2000} and for a totally asymmetric version of the model~\cite{krug_g2000} defined on infinite line 
with different dynamics. 
For parallel updating scheme a mean-field calculation has been shown to be exact for certain parameters~\cite{zielen_s2002a} 
and a Matrix Product Ansatz has been developed~\cite{zielen_s2003}. Finally, it has been shown that a condensation transition occurs 
in the related mass transfer model if one imposes a cutoff on the transferred mass 
( \emph{i.e} on the amount of jump  made by the particles in RAP picture)~\cite{zielen_s2002b}. 

Recently there has been a considerable interest in studying the motion of a special driven tagged particle in presence of other non-driven interacting particles. Such 
special particle is called the tracer particle. In experimental studies, 
driven tracers in quiescent media have been used to probe rheological properties of complex media 
such as DNA~\cite{gutsche2008}, polymers~\cite{kruger_r2009}, granular media~\cite{candelier_d2010, pesic2012} or colloidal crystals~\cite{dullens_b2011}.
Some practical examples of biased tracer are :
charged impurity being driven by applied electric field or a colloidal particle being pulled by optical tweezer in presence of other colloid particles 
performing random motion. On the theoretical side, problems with driven tracer have been studied in the context of SEP where both particle number 
conservation and non-conservation (absorption/desorption) have been considered ~\cite{burlatsky1992, burlatsky1996, landim_o_v1998, benichou1999}.
In presence of a driven tracer on an infinite line, the particle density profile is inhomogeneous around the tracer and in absence of 
absorption or desorption, the current flowing across the system vanishes in the large time limit, as does the velocity of the tracer 
\cite{burlatsky1992, burlatsky1996, landim_o_v1998}.
Other quantities like the mean and the variance of the position of the tracer have also been studied \cite{Beni_et_al2013}. 
In this contribution we derive new results for RAP concerning the statistics of the gaps between particles on a ring in presence of a 
tracer particle which may be driven.

More precisely we consider $N$ particles moving according to RAP on a ring of size $L$. In a small time duration $dt$, any particle, except the tracer particle, 
jumps to the right or left by a random fraction of the space available up to the neighboring particle on the right or left, respectively, with equal probability $dt/2$.
On the other hand the tracer particle moves to the right or left by a 
random fraction of the space available in the respective direction with probability $pdt$ and $qdt$, see fig.~\ref{fig:scheme}.
In this paper we study the statistics of the gaps between neighboring particles in the stationary state (SS) for the following two cases : (i) $p=q$ (ii) $p\neq q$. 
Note that the dynamics does not satisfy detailed balance even in the $p=q$ case, hence keeps the system always out of equilibrium.
Using a mapping from RAP model of particles to an equivalent (except for a global translation) MT model, we find various interesting 
exact results related to gap statistics. 
The summary of these results are given below :
\begin{itemize}
 \item When $p=q$ \emph{i.e.} the tracer particle is not driven, all the particles are moving symmetrically. 
In this case, for a large class of jump distributions $R(\eta)$ that satisfy some necessary and sufficient condition, we find that the stationary joint distribution 
of the gaps $\{g_i,~i=1,2,...,N\}$ takes the following universal form :   
\begin{equation}
\label{eq:statbeta}
 P_{N,L}(g_1,g_2,...,g_N) = \frac{1}{\cZ_{N,L}(\beta)} \prod_{i=1}^{N} g_i^{\beta-1} ~~\delta \left(\sum_{i=1}^{N} g_i - L \right ) ,
\end{equation}
where $\cZ_{N,L}(\beta)$ is the normalization constant and the delta function represents the global constraint due to total mass conservation in the MT picture. 
The parameter $\beta > 0$ is given by $\beta=\frac{\mu_1}{\mu_2}-1$, where $\mu_k=\int_0^1\eta^kR(\eta)d\eta$ is the $k^{\text{th}}$ 
moment of the jump distribution $R(\eta)$. Similar factorized joint distributions with power law weight functions have also been obtained in the 
context of the q-model of force fluctuations in granular media \cite{coppersmith1996} as well as in a totally asymmetric version of the RAP on an infinite 
line with parallel dynamics \cite{zielen_s2002a}.
For arbitrary jump distribution $R(\eta)$ we find that the average mass profile is given by 
$\langle g_i\rangle=\omega_0=L/N$ and the two point gap-gap (mass-mass) correlation 
$d_{i,j} = \langle g_{i}g_{j}\rangle-\langle g_{i}\rangle \langle g_{j}\rangle $ is expressed in terms of $\mu_1$ and $\mu_2$ as
\begin{equation}
 d_{i,j} = \left( \frac{\mu_2 \omega_0^2}{\mu_1+(N-1)(\mu_1-\mu_2)}\right) [N\delta_{i,j}-1].
\end{equation}

\item In the second case when $p\neq q$ \emph{i.e.} the tracer particle is driven, the factorization 
form of the joint distribution of the gaps does not hold. In this case the stationary state has 
a global current associated with the non-zero mean velocity of the tracer, which supports an inhomogeneous mean mass profile in the MT picture. 
We compute this average mass profile $m_i=\langle g_i \rangle$ and it is given by 
\begin{equation}
 m_i= \frac{\omega_0}{(p+q)(N-1)+1}~\left[2(p-q)i+(2qN-2p+1)\right]. 
\end{equation}
We also compute the pair correlation $d_{i,j}=\langle g_i g_j\rangle-\langle g_i \rangle \langle g_j \rangle$ in the steady state. 
In the thermodynamic limit \emph{i.e.} for both, $L \to \infty$ and $N \to \infty$ limit, while keeping the mean gap density $\omega_0=L/N$ fixed, 
we find that the average mass $m_i$, 
variance $d_{i,i}$ and the correlation $d_{i,j}$ scale as
\begin{eqnarray}
 m_i &=& \omega_0~\mathcal{M}\left ( \frac{i}{N}\right) +o(1/N), \\
 d_{i,j} &=& \omega_0^2 ~\frac{1}{N}~\mathcal{D}\left ( \frac{i}{N},~\frac{j}{N}\right)+o(1/N), \qquad i \neq j \\
 d_{i,i} &=& \omega_0^2 \left[ \mathcal{C}_0\left ( \frac{i}{N}\right) + \frac{1}{N}~\mathcal{C}_1\left ( \frac{i}{N}\right) 
+ \frac{1}{N}~\mathcal{D}\left ( \frac{i}{N},~\frac{i}{N}\right) +o(1/N)\right],
\end{eqnarray}
where $o(\ell)$ represents order smaller that $\ell$. We find explicit expressions of the scaling functions $\mathcal{M}(x)$, 
$\mathcal{C}_0(x)$ and $\mathcal{C}_1(x)$ in \eqref{msqsclng}, \eqref{eq:C0} and \eqref{eq:C1}, respectively. The explicit expression 
of $\mathcal{D}(x,y)$ is given in \eqref{sol-Dxy} for $q=0$ whereas the expression for general $q$ can be obtained following the analysis 
given in Appendix \ref{derivation-Green-func}. 

\end{itemize}

\noindent
The paper is organized as follows. In section~\ref{section:model} the model and basic equations are written. We then study the RAP without driven tracer 
in section~\ref{section:srap}, where we focus on factorized stationary distributions with a global constraint and on the jump distributions for which 
they occur. In the next section~\ref{section:tsrap} we study the influence of a driven tracer. For this case, we obtain single site gap distribution, mean gap and 
two-point correlations of the gaps in the thermodynamic limit. Finally, section~\ref{section:ccl} concludes the paper. 
Some details are given in the appendices \ref{appendix-phihat_s}, \ref{derivation-sclng-diff-eq} and \ref{derivation-Green-func}.

\section{Model and basic equations}
\label{section:model}

\begin{figure}[!ht]
	\begin{center}
		\includegraphics[width=0.7\textwidth]{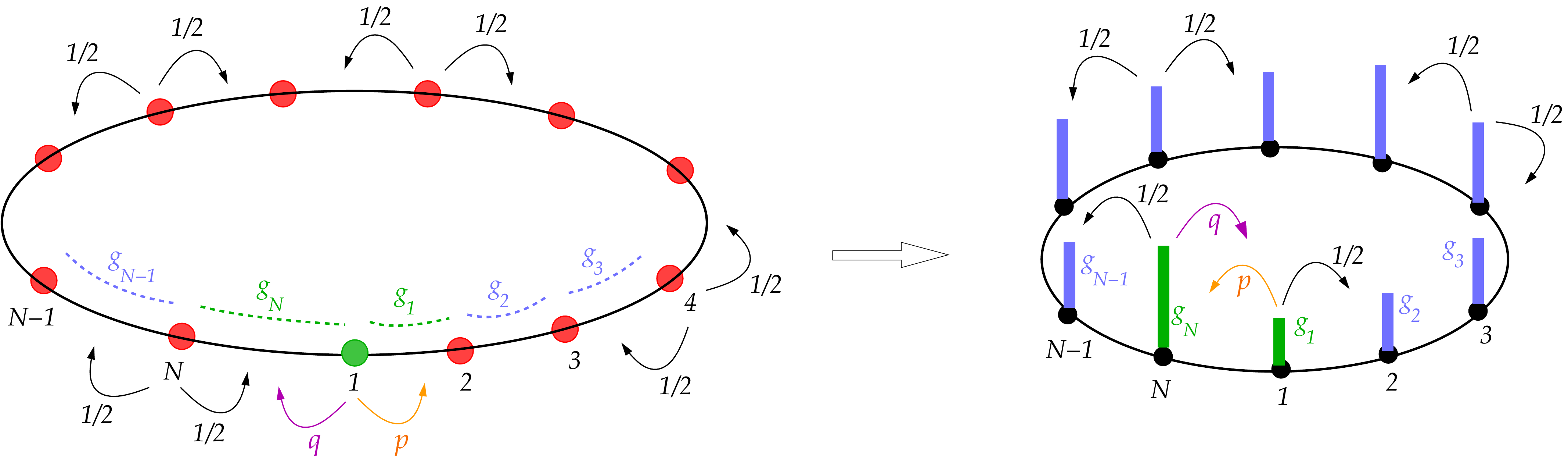} 
	\end{center}
	\caption{\small Schematic diagram of the RAP model with $N$ particles and a driven tracer particle (particle $\#$1 ) is presented on the left 
                 and the diagram of the corresponding mass model is presented on the right. 
	Each particle of the RAP can be mapped onto a link between two lattice sites in the mass model, the tracer particle corresponds to a particular bond 
        (between sites $N$ and $1$). When going from the RAP to the mass model one loses information about a translational degree of freedom. }  
	\label{fig:scheme}
\end{figure}

We consider $N$ particles moving on a ring of size $L$ ( see Fig.\ref{fig:scheme}) which are labeled as $i=1,\ldots,N$. Among these $N$ particles, let us consider
the first particle as the tracer particle which may be driven. Their positions at time $t$ 
are denoted by $x_i(t)$ for $i=1,\ldots,N$.
The dynamics of the particles are given as follows. 
In an infinitesimal time interval $t$ to $t+dt$, any particle (say $i^{\text{th}}$) other than the tracer, jumps from $x_i(t)$, 
either to the right or to the left with probability $dt/2$  
and with probability $(1-dt)$ it stays at $x_i(t)$. The tracer particle jumps from $x_{1}(t)$ to the right with probability $pdt$, to the left with probability $qdt$ 
and does not jump with probability $(1-(p+q)dt)$.
The jump length, either to the right or to the left, made by any particle is a random fraction 
of the space available between the particle and its neighboring particle to the right or to the left.
For example, the $i^{\text{th}}$ particle jumps by an amount $\eta_i^r [x_{i+1}(t) - x_i(t)]$ to the right and 
by an amount $\eta_i^l [x_{i-1}(t) - x_i(t)]$ to the left. The random variables $\eta_i^{r,l}$ are independently chosen from the interval $[0,~1]$ and each is distributed 
according to the same distribution $R(\eta)$. For most of our calculations in this paper we consider arbitrary $R(\eta)$, unless it is specified.
Following the dynamics of the particles for the homogeneous case from \cite{rajesh_m2001}, 
one can write down the stochastic evolution equations for the locations of the particles as: 
\begin{eqnarray}
x_i(t+dt)&=& 
\begin{cases}
& x_i(t) + \eta_i^r(x_{i+1}(t)-x_i(t)),~~\text{with~Prob.~}~R(\eta_i^r)d\eta_i^r~\frac{dt}{2}, \\
& x_i(t) + \eta_i^l(x_{i-1}(t)-x_i(t)),~~\text{with~Prob.~}~R(\eta_i^l)d\eta_i^l~\frac{dt}{2},~~~~~i \neq 1,\\
& x_i(t),~~~~~~~~~~~~~~~~~~~~~~~~~~~~~~\text{with~Prob.~}~~~1-dt,
\end{cases} \label{dyna-ith}  \\
\text{and} && \nonumber \\
x_{1}(t+dt)&=& 
\begin{cases}
& x_{1}(t) + \eta_{1}^r(x_{2}(t)-x_{1}(t)),~~~\text{with~Prob.~}~R(\eta_{1}^r)d\eta_{1}^r~pdt, \\
& x_{1}(t) + \eta_{1}^l(x_{0}(t)-x_{1}(t)),~~~\text{with~Prob.~}~R(\eta_{1}^l)d\eta_{1}^l~qdt,\\
& x_{1}(t),~~~~~~~~~~~~~~~~~~~~~~~~~~~~~\text{with~Prob.~}~1-(p+q)dt,
\end{cases} \label{dyna-tracer}
\end{eqnarray}
where to impose the periodicity in the problem, we introduced two auxiliary particles with positions $x_{N+1}(t) \equiv x_1(t) + L$ 
and $x_0(t) \equiv x_N(t) -L$ for all $t$. Note that the dynamics is invariant under simultaneous dilation of $L$ and of all the positions. In this paper, 
the results are however stated for general $L$.

Since we are interested in the statistics of the gaps $g_i=x_{i+1}-x_i,~~i=1,2,...,N$, it is convenient 
to work with an equivalent and appropriate mass transfer (MT) model with $N$ sites. 
The mass transfer model is defined as follows. Corresponding to the $N$ particles in the RAP, we consider 
a periodic one dimensional lattice of $N$ sites with a mass at each site. Particles from the RAP picture are mapped 
to the links between lattice sites in the MT picture. 
For example, the $i^{\text{th}}$ particle in RAP corresponds to the link between sites $(i-1)$ and  $i$ in MT model 
whereas the mass at site $i$ is equal to the gap $g_i=x_{i+1}-x_i$ between the $i^{\text{th}}$ and $(i+1)^{\text{th}}$ particle in RAP. 
As a result for each configuration 
${\bf X}(t)=[~x_1(t), x_2(t),\ldots, x_N(t)~]$ of the positions of the particles in the RAP 
we have a unique mass configuration ${\bf G}(t) = [~g_1(t),g_2(t), \ldots, g_{N}(t)~]$ in the MT model 
where $g_i=x_{i+1}-x_i,~~i=1,2,...,N$. The opposite is however not true, as the instantaneous 
configuration of the masses leaves freedom for a global translation of the system in the RAP. 
Such mapping from particle model to mass model have been considered in other contexts like from the exclusion process
to the zero range process~\cite{evans_h2005}.

Concerning the dynamics, a hop of the $i^{\text{th}}$ particle by $\eta(x_{i+1}-x_i)$ towards the $(i+1)^{\text{th}}$ particle in the RAP corresponds 
to a transfer of mass $\eta g_i\equiv\eta(x_{i+1}-x_i) $ from $i^{\text{th}}$ site to the $(i-1)^{\text{th}}$ site in the MT model whereas 
a hop of the $i^{\text{th}}$ particle by $\eta(x_{i}-x_{i-1})$ towards the $(i-1)^{\text{th}}$ particle in the RAP corresponds 
to a transfer of mass $\eta g_{i-1}\equiv\eta(x_{i}-x_{i-1}) $ from $(i-1)^{\text{th}}$ site to the $i^{\text{th}}$ site. 
More precisely, the updating rules for the configurations in the MT model are given by :
\begin{equation}
\label{dyna-mass}
g_i(t+dt)=g_i(t)+\sigma^{i+1}_r \eta_{i+1} g_{i+1}(t)+\sigma^{i-1}_l \eta_{i-1} g_{i-1}(t)-(\sigma^{i}_l +\sigma^{i}_r)~\eta_{i} g_{i}(t),
\end{equation}
where the $\eta$ variables are independent and identically distributed according to $R(\eta)$ 
and $\sigma_{r,l}^i$ are $1$ with probability $\frac{dt}{2}$ and $0$ otherwise except  
for $\sigma_{r}^N$ and $\sigma_{l}^1$. The random variable $\sigma_{r}^N$ is $1$ with probability $q dt$ and $0$ with probability $1-q dt$. 
Similarly, $\sigma_{l}^1$ is $1$ with probability $p dt$ and $0$ with probability $1-p dt$. 
The periodicity is imposed by $g_{N+1}(t)= g_1(t)$ and $g_{0}(t)= g_N(t)$, for all $t$. 

\vspace{0.5cm}
\noindent
{\bf Steady State Master Equation :} The dynamics described in  (\ref{dyna-mass}) will take the system eventually to a stationary state. 
If $P_{N,L}({\bf G})$ represents the steady state joint probability distribution of the configuration ${\bf G} = (g_1,g_2, \ldots, g_{N})$, 
then it satisfies the following master equation
\begin{eqnarray}
\label{eq:meqsrap}
&&(N+p+q-1) P_{N,L}({\bf G})= \sum_{i=1}^{N} \int_{0}^\infty dg_{i}' \int_{0}^\infty d g_{i+1}'\int_{0}^1 d \eta ~R(\eta)~ P_{N,L}({\bf G}_{i,i+1}') \nonumber \\
&&~~~~~~~\times~\big{[}(~1/2+(q-1/2)\delta_{i,N})~\delta(g_i - g_i' + \eta g_i')~ \delta(g_{i+1} - g_{i+1}' - \eta g_{i}')   \label{eq:meqss}   \\
&&~~~~~~~~~~~~~+~(~1/2+(p-1/2)\delta_{i,N})~\delta(g_i - g_i' - \eta g_{i+1}')~\delta(g_{i+1} - g_{i+1}' + \eta g_{i+1}') \big{]}, \nonumber
\end{eqnarray}
where we used the shorthand notation ${\bf G}_{i,i+1}' = (g_1, \ldots, g_{i-1}, g_i', g_{i+1}', g_{i+2}, \ldots, g_{N})$. 
The conservation of the number of particles follows naturally from the fact that the lattice size is fixed. Moreover, 
the dynamics clearly conserves the total mass \emph{i.e.} we have $\sum_{i=1}^{N} g_i(t) = L$, at all times. As stated in the introduction we are interested in 
the statistics of the gaps/masses $g_i=x_{i+1}-x_i$ in SS for the following two cases : (i) $p=q$ and (ii) $p\neq q$. In the next section we analyze the first case.

\section{Unbiased Tracer Case : $p=q$ }
\label{section:srap}
In this case the tracer particle is not driven and all the particles are moving symmetrically. In the large time limit, the inter-particle spacings or gaps will 
reach a stationary state. To understand completely the statistics of the gaps in SS, 
one would require to find the joint probability distribution function (JPDF) of all the gaps $g_i;~i=1,2,...,N$. Clearly, 
this JPDF could be different for different choices of jump distributions $R(\eta)$. Finding this JPDF for arbitrary $R(\eta)$ is generally a hard task. 
We ask a relatively simpler question : for what choices of the jump distribution does the stationary JPDF have a factorized form 
\begin{equation}
\label{eq:srapstatansatz}
P_{N,L}({\bf G}) = \frac{1}{\cZ_{N,L}} \prod_{i=1}^{N} w_i(g_i)~ \delta \left(\sum_{i=1}^{N} g_i - L \right) = \frac{W_{N,L}({\bf G})}{\cZ_{N,L}} ,
\end{equation}
except for a global mass conservation condition represented by the delta function. Here 
$\cZ_{N,L} = \int_{{\bf G} \in [0;\infty)^N} W_{N,L}({\bf G}) d {\bf G}$ ensures normalization.
Analogous questions have also been asked in different other contexts, for example in mass transport models \cite{krug_g2000, rajesh_m2000, zielen_s2002a}, 
in zero range processes \cite{Evans_M_Z2004, zia_e_m2004} and in finite range processes \cite{Cha_P_M2015}.
In the following subsection we will see that there exists a large class of $R(\eta)$ for which the above form for JPDF is true. Moreover we will see that the 
weights $w_i(g)$ corresponding to individual sites are $i$ independent and the weight functions have universal form.

\subsection{Determination of the weight functions $w_i(g)$}
\label{subsection:srapmassstat}
\noindent
To obtain the weight functions $w_i(g)$, we first insert $P_{N,L}({\bf G})$ from  (\ref{eq:srapstatansatz}) in the SS master equation (\ref{eq:meqsrap}) and then 
integrate over all $g_j$s except $j=i$. We get 
\begin{eqnarray}
\label{eq:stat1mass}
&&2[1+(p-1/2)(\delta_{i,1}+\delta_{i,N})] ~W_{N,L}^{(1)}(g_i)  \\
&&~~~~~~~~~= \int_{0}^{\infty} d g_{i-1} \int_0^{\infty}d g_i' \int_{0}^{1} d \eta ~R(\eta)~ [1/2+(p-1/2)\delta_{i,1}]\left[\delta(g_i -g_i' + \eta g_i') 
+ \delta(g_i -g_i' - \eta g_{i-1})\right]~ W_{N,L}^{(2)}(g_{i-1},g_i') \nonumber \\
&&~~~~~~~~~+ \int_{0}^{\infty} d g_{i+1} \int_0^{\infty}d g_i' \int_{0}^{1} d \eta ~R(\eta) ~[1/2+(p-1/2)\delta_{i,N}]\left[ \delta(g_i -g_i' + \eta g_i') 
+ \delta(g_i -g_i' - \eta g_{i+1})\right]~ W_{N,L}^{(2)}(g_i',g_{i+1}), \nonumber
\end{eqnarray}
where 
\begin{eqnarray}
W_{N,L}^{(1)}(g_i) = \left( \prod_{j\neq i}^{N}\int_0^{\infty} dg_j~\right) W_{N,L}({\bf G}),~~\text{and}~~
W_{N,L}^{(2)}(g_i,g_k) = \left( \prod_{j\neq i \neq k}^{N}\int_0^{\infty} dg_j~\right) W_{N,L}({\bf G}). \label{marginalP_1} 
\end{eqnarray}
We define the Laplace transform of any function $f(x)$ as $\hat{f}(s)=\int_0^{\infty}f(x)\ee^{-sx}~dx$ where $s$ is the Laplace conjugate of $x$. 
Taking Laplace transform over $L$ as well as over $g_i$ on both sides of  (\ref{eq:stat1mass}) we get
\begin{eqnarray}
\label{eq:stat1masslt}
2 \frac{\hw_i(s+s')}{\hw_i(s)} &=& \int_{\eta=0}^{1} d \eta ~R(\eta) \frac{\hw_i(s+(1-\eta) s')}{\hw_i(s)} 
+ \frac{\hw_i(s+s')}{\hw_i(s)} \int_{\eta=0}^{1} d \eta~ R(\eta) 
\Bigg{[}\frac{1/2+(p-1/2)\delta_{i,1}}{1+(p-1/2)(\delta_{i,1}+\delta_{i,N})}\frac{\hw_{i-1}(s+\eta s')}{\hw_{i-1}(s)} \nonumber \\
&&~~~~~~~~~~~~~~~~~~~~~~~~~~+ \frac{1/2+(p-1/2)\delta_{i,N}}{1+(p-1/2)(\delta_{i,1}+\delta_{i,N})}\frac{\hw_{i+1}(s+\eta s')}{\hw_{i+1}(s)} \Bigg{]},
\end{eqnarray}
where $s$ is Laplace conjugate to $L$ and $s'$ is Laplace conjugate to $g_i$. While deriving the above equation we have assumed that the Laplace transform of 
$w_i(g)$, defined by $\hw_i(s) = \int_{0}^\infty dg~\ee^{-s g} w_i(g)$, exists. 
Equation (\ref{eq:stat1masslt}) provides the condition satisfied by $w_i(g_i)$, in order to get a factorized JPDF as in  (\ref{eq:srapstatansatz}). 
To find the solution for $\hw_i(s)$, let us expand both sides of  (\ref{eq:stat1masslt}) in powers of $s'$ and equate coefficients of each power on both sides. 
One can easily see that at order $s'^0$,   (\ref{eq:stat1masslt}) is automatically satisfied because of the normalization $\int_0^1d\eta R(\eta)=1$. 
At order $s'^1$, we get
\begin{equation}
 2\frac{\hw_i'(s)}{\hw_i(s)} = \left( \frac{\hw_{i-1}'(s)}{\hw_{i-1}(s)} + \frac{\hw_{i+1}'(s)}{\hw_{i+1}(s)}\right) 
+ (2p-1)(\delta_{i,1}-\delta_{i,N})\left( \frac{\hw_{N}'(s)}{\hw_{N}(s)} - \frac{\hw_{1}'(s)}{\hw_{1}(s)}\right).
\end{equation}
The general solution of the above equation is $\frac{\hw_i'(s)}{\hw_i(s)}= A(s)~i~+~B(s)$ where $A(s)$ and $B(s)$ are $s$ dependent constants. 
From the boundary conditions at $i=1$ and $i=N$, we determine $A(s)=0$. Hence we find that $\hw_i(s)$ is independent of site index $i$. 
To proceed further, we now look at the expansion of  (\ref{eq:stat1masslt}) 
at order $s'^2$ and get
\begin{equation}
\label{eq:sp2}
(\mu_2-\mu_1) \frac{\hw''(s)}{\hw(s)} + \mu_1 \left( \frac{\hw'(s)}{\hw(s)} \right)^2 = 0.
\end{equation}
The above equation can be easily solved to get 
\begin{equation}
\label{eq:beta}
\hw(s) = A_0(s+B_0)^{-\beta},~~~\text{where},~~~\beta = \frac{\mu_1-\mu_2}{\mu_2},
\end{equation}
and $A_0$, $B_0$ are constants. Taking inverse Laplace transform we get $w(g) = \frac{A_0g^{\beta-1}e^{-B_0g}}{\Gamma[\beta]}$ where $\Gamma[x]$ is the Gamma function.
Inserting this form of $w(g)$ in  (\ref{eq:srapstatansatz}) and absorbing all the constants in the normalization constant $\cZ_{N,L}$ we arrive at the result 
\begin{equation}
\label{eq:statbeta1}
 P_{N,L}(g_1,g_2,...,g_N) = \frac{1}{\cZ_{N,L}(\beta)} \prod_{i=1}^{N} g_i^{\beta-1} ~~\delta \left(\sum_{i=1}^{N} g_i - L \right ) ,
\end{equation}
as stated in  (\ref{eq:statbeta}). 
The normalization constant $\cZ_{N,L}$ can be computed to give
\begin{equation}
\label{eq:ZNL}
\cZ_{N,L} = \cZ_{N,L}(\beta) = \frac{L^{\beta N - 1} \Gamma[\beta]^N}{\Gamma[\beta N]}.
\end{equation}

\subsection{Jumping distributions that yield  (\ref{eq:statbeta1})}
\label{Jump-dist}
\noindent
In the previous section we have seen that if there exists a jump distribution $R(\eta)$ for which JPDF $P_{N,L}({\bf G})$ is in the form (\ref{eq:srapstatansatz}) then, 
$w_i(g_i)$ should always be equal to $g_i^{\beta-1}$ for all $i$ where $\beta=\frac{\mu_1-\mu_2}{\mu_2}$.
Now we assume  (\ref{eq:statbeta1}) to be true and find what conditions $R(\eta)$ should satisfy. To get that condition 
we start with \eqref{eq:stat1mass} which implies \eqref{eq:stat1masslt}. Using now $\int_0^{\infty}dg~e^{-sg}~g^{\beta-1}=\Gamma[\beta]s^{-\beta}$ 
in \eqref{eq:stat1masslt} we find, 
\begin{equation}
\label{eq:stat1massltu}
\frac{2}{(1+u)^\beta} = \int_{0}^{1} d \eta ~\frac{R(\eta)}{(1+(1-\eta) u)^\beta} 
+ \frac{1}{(1+u)^\beta} \int_{0}^{1} d \eta ~\frac{R(\eta)}{(1+\eta u)^\beta},
\end{equation}
for all $u=(s'/s) \geq 0$. This is a necessary condition that $R(\eta)$ should satisfy to get a factorized JPDF in SS. 
For distributions $R(\eta)$ satisfying \eqref{eq:stat1massltu} such that the system is ergodic, we expect the stationary state to be unique. Consequently the condition 
\eqref{eq:stat1massltu} is sufficient to ensure that the steady state joint gap distribution is given by \eqref{eq:statbeta1}. 
Analogous conditions satisfied by hopping rates, have been obtained in other mass transport models \cite{Evans_M_Z2004, majumdar2010} 
and in finite range process \cite{Cha_P_M2015}. 

\noindent
We next find some solutions of  (\ref{eq:stat1massltu}) for particular values of $\beta$, as well as a family of solutions that covers the whole $\beta$ range. 

\begin{itemize}
 \item 
Let us first consider the $\beta=0$ case. Formally, $\beta=0$ corresponds to 
$\mu_1 = \mu_2$, \textit{i.e.} $R(\eta) = \delta(\eta-1)$. For this choice of $R(\eta)$ system breaks ergodicity and the stationary state is a 
trivial one with all the particles at a single point. As a result, $P_{N,L}({\bf G})=(1/N)\sum_{j=1}^N\prod_{i\neq j}^{N} \delta(g_i)~\delta(\sum_kg_k-L)$.

\item
A very special case is $\beta=1$, for which the stationary distribution~~\eqref{eq:statbeta1} is uniform over the allowed configurations. 
For this case some simple examples of the solutions are $R(\eta) = \delta(\eta-\frac{1}{2})$, 
$R(\eta) = \frac{\Gamma[1+2\alpha]}{\Gamma[\alpha]\Gamma[\alpha+1]}\eta^{\alpha-1} (1-\eta)^\alpha~;~\alpha \geq 0$ 
which can be easily verified by directly inserting them in (\ref{eq:stat1massltu}) with $\beta=1$. 
Interestingly, one can find \emph{all} the possible solutions of~\eqref{eq:stat1massltu} 
in this particular case. 
Defining $v = \frac{u^2}{1+u}$ and $R(\eta) = (1-\eta) f(\eta)$, elementary manipulations show that for $\beta=1$, eq.~\eqref{eq:stat1massltu} is equivalent to
\begin{equation}
\label{eq:beta1}
 \int_{0}^{1} d \eta~\frac{(1-\eta) \eta^2 (2 \eta -1)}{1+v \eta (1-\eta)} [f(\eta) - f(1-\eta)] =0, \qquad \forall~v > 0.
\end{equation}
The factor multiplying $f(\eta) - f(1-\eta)$ on the RHS is always positive, which implies that $f$ is symmetric with respect 
to $\frac{1}{2}$. Hence all possible solutions are of the form $R(\eta) = (1-\eta) f(\eta)$ with $f(\eta) = f(1-\eta)$ 
for all $\eta \in (0,1)$. The converse is shown to be true iff the integral of $f$ is equal to $2$. Therefore, the set of solutions for $\beta=1$ are
\begin{equation}
\label{eq:Retabeta1}
R(\eta) = (1-\eta) f(\eta),~~\text{with},~~f(\eta)=f(1-\eta),~~\text{and}~~\int_{0}^{1} f(\eta) ~d \eta = 2.
\end{equation}

\item
General $\beta$ : One can find a simple family of solutions that spans the whole range of $\beta>0$ values,
\begin{equation}
\label{eq:Retabetaany}
R(\eta) = 2 \beta (1-\eta)^{2 \beta -1},~\beta > 0.
\end{equation}
In particular, taking $\beta=\frac{1}{2}$ gives the uniform distribution. 

\item
The situation is quite interesting for $\beta\rightarrow\infty$. One can check that the following scaling form 
 $R(\eta) = \beta \ee^{-\beta \eta} \phi(\beta \eta)$ solves ~\eqref{eq:stat1massltu} for $\beta\rightarrow\infty$,
 where $\phi(y)$ is a real, positive function for $y\geq 0$ and its Laplace transform $\hat{\phi}$ satisfies 
\begin{equation}
\label{eq:stat1masslty}
\int_{0}^\infty d y~\phi(y)~ \left(\ee^{-y (1+u)} + \ee^{-\frac{y}{1+u}} \right)  = \hat{\phi}(1+u) + \hat{\phi}\left(\frac{1}{1+u}\right)=2,~~~\text{for}~~~u \geq 0.
\end{equation}
Note that $\hat{\phi}(1)=1$. There exists an infinite number of solutions of the above equation. 
For example, one class of solution can be chosen as $\hat{\phi}(s)=\frac{H_u(s)}{H_b(s)}$ where, both $H_u(s)$ and $H_b(s)$, are polynomials of same order, say $m$. 
There are $2(m+1)$ constants corresponding to $m$ coefficients of different powers of $s$ associated to the two polynomials among which only $m+1$ are independent, 
as they are linked by  (\ref{eq:stat1masslty}). 
Now these $m+1$ coefficients have to be chosen such that $\hat{\phi}(1)=1$ and $\phi(y)$ is real and positive for $y \geq 0$. 
The simplest of them is $\hat{\phi}(s) = \frac{2}{s+1}$ which gives the solution
\begin{equation}
\label{eq:Retabetainf}
R(\eta) = 2 \beta \ee^{-2 \beta \eta} ~~~\text{for}~~~  \beta \rightarrow \infty.
\end{equation}
Some other examples of solutions in $\hat{\phi}(s)=\frac{H_u(s)}{H_b(s)}$ form are given in appendix (\ref{appendix-phihat_s}). 
\end{itemize}
The above analysis suggests that there are possibly infinitely many solutions of the  (\ref{eq:stat1massltu}) for any $\beta >0$. Although we found all the solutions 
for $\beta=0$ and $\beta=1$, and a somewhat simpler characterization of them for large $\beta$, a proper characterization of all the possible solutions for
 arbitrary $\beta$ seems difficult. 
We are however able to make a few general remarks about the properties of the solutions of \,\eqref{eq:stat1massltu}.
\begin{itemize}
	\item 
	If $R_1(\eta)$ and $R_2(\eta)$ both are solutions of  (\ref{eq:stat1massltu}) corresponding to the same $\beta$, then any 
	normalized linear combination of them \emph{i.e.} $\bar{R}(\eta)=r R_1(\eta)+(1-r)R_2(\eta)$ for $0\leq r \leq 1$ is also a solution. 
	\item
	Secondly, the system is invariant by dilation of time. Replacing $t$ by $\frac{t}{1-r}$ is equivalant to replacing $R(\eta)$ 
        by $r \delta(\eta) + (1-r) R(\eta)$ for $0 \leq r \leq 1$. Indeed, with this latter choice a chosen particle hops with probability $1-r$ only, 
        and nothing happens with probability $r$. 
	Clearly, the values of $\mu_1$ and $\mu_2$ depend on the value of $r$ however $r$ should not appear in the stationary distribution. 
	The stationary distribution therefore cannot depend on $\mu_1$ and $\mu_2$ independently, but only through a particular combination in which $r$ cancels,
	\emph{i.e} $\beta=\frac{\mu_1-\mu_2}{\mu_2}$. All the solutions given above are understood up to a dilation of time.
\end{itemize}

\noindent
Equation~\eqref{eq:stat1massltu} necessary for getting the stationary state \eqref{eq:statbeta1} is quite robust, 
as the same equation can be obtained in the case where all the particles are 
identical but asymmetrically moving. In a much more general case where each particle has its own hopping rates $p_i+r$ and $p_i-r$ 
to the right and to the left, respectively, it can be shown that one still obtains \eqref{eq:stat1massltu} as the condition 
for getting the factorized SS  \eqref{eq:statbeta1}.
But when all the particles have arbitrary jumping rates, possibly there is no jump distribution $R(\eta)$ available for which 
such factorized states exists and this most general case has not been studied in the literature. 

\subsection{Single mass distribution}
\label{single-mass-dist-p-eq-q}
\begin{figure}[!ht]
	\begin{center}
		\includegraphics[width=0.6\textwidth]{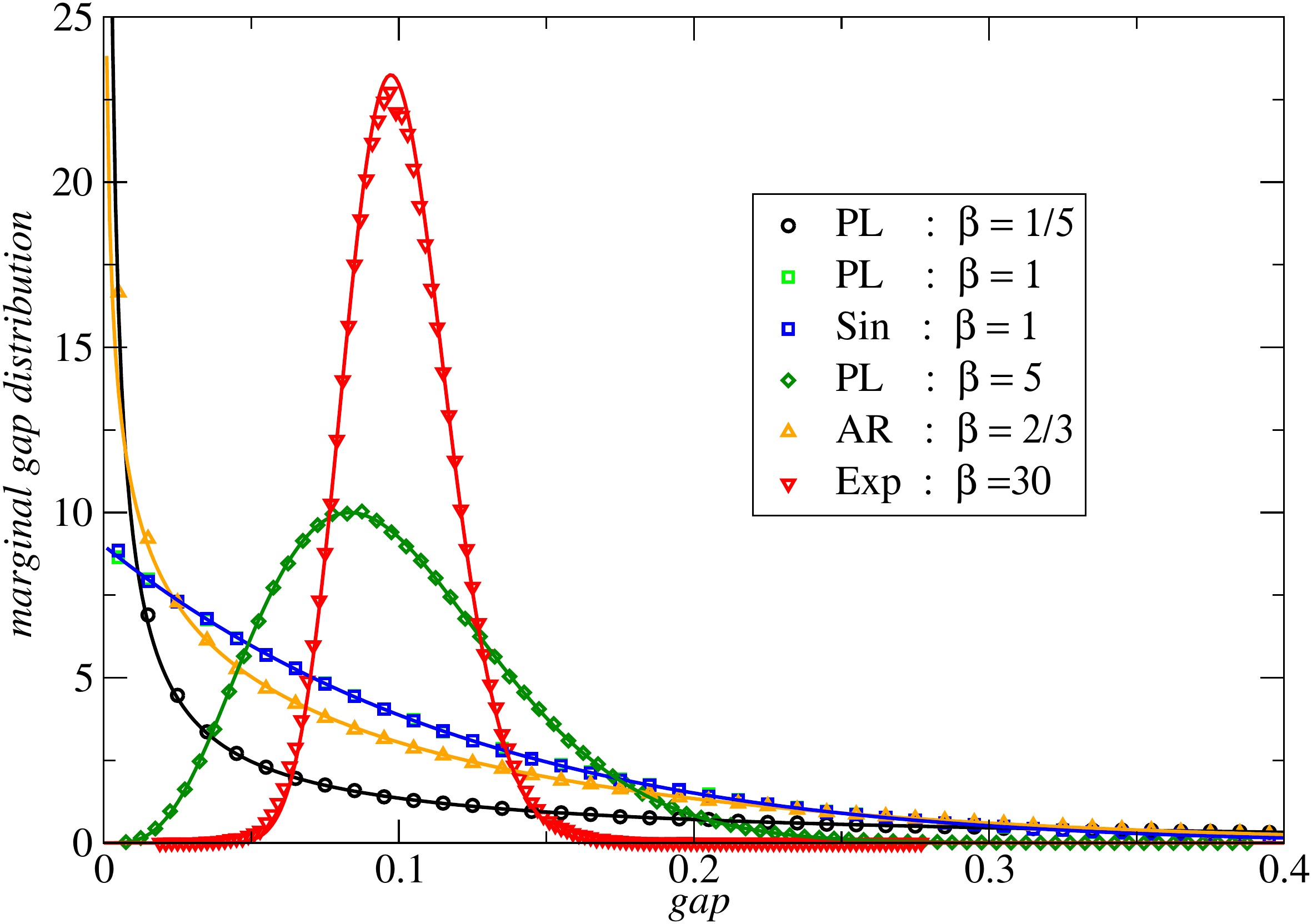} 
	\end{center}
	\caption{\small Stationary marginal distribution of a the gaps/masses in the $p=q$ case for $N=10$, $L=1$. 
In all plots, the symbols represent data obtained from numerical simulation and the corresponding solid lines are obtained from 
the theoretical expression in \,~\eqref{eq:marginalM}. Different curves correspond to different forms of jump distribution $R(\eta)$. 
Here 'PL' corresponds to power law : $R(\eta)=2\beta (1-\eta)^{2\beta-1}$ for $\beta=0.2, 1.0$
and $5.0$. 'Sin' represents sin law $R(\eta)=\pi (1-\eta) \sin(\pi \eta)$ for which $\beta=1$. 
'Exp' represents exponential law : $R(\eta)=2 \beta e^{-2\beta \eta}$ 
for $\beta=30$. Finally, 'AR' corresponds to $R(\eta)=6\eta (1-\eta)$ for which $\beta=2/3$. Although the distribution form 'AR' does not satisfy 
 (\ref{eq:stat1massltu}), we see that the numerically obtained marginal distribution still matches with 
the analytical expression in \,~\eqref{eq:marginalM} with $\beta=2/3$. }  
\label{fig:p1msrap}
\end{figure}
\noindent
In this subsection we compute the marginal mass distribution and compare it to numerically obtained distributions for different jump distributions $R(\eta)$.
The marginal distribution $P_{N,L}({g_i})$ can be readily computed form $P_{N,L}({\bf G})$ by integrating out all $g_j$s except $g_i$. 
For factorized stationary states we get from  (\ref{eq:statbeta1}) that
\begin{equation}
\label{eq:marginalM}
P_{N,L} (g) = \frac{1}{L} \frac{\Gamma[\beta N]}{\Gamma[\beta] \Gamma[\beta (N-1)]}
 \left( \frac{g}{L}\right)^{\beta-1} \left( 1 - \frac{g}{L}\right)^{\beta N - \beta -1},~~~0  \leq g \leq L,~~1\leq i \leq N.
\end{equation}
The corresponding moments of the above distribution for arbitrary $N$ and $L$ are 
\begin{equation}
\label{eq:momentsstatsrap}
\langle g^k \rangle = L^k \frac{\Gamma \left[ \beta+k\right] \Gamma \left[\beta N\right]}{\Gamma[\beta] \Gamma \left[ \beta N + k\right]},~~~\forall k \geq 0.
\end{equation}
In the limit $L \rightarrow \infty$ and $N \to \infty$ with fixed density $\omega_0 = \frac{L}{N}$ the marginal mass distribution in (\ref{eq:marginalM}) becomes 
\begin{equation}
\label{eq:marginalMLinf}
\lim_{N \rightarrow \infty} P_{\omega_0 N,N} (g) = \frac{\beta^\beta}{\omega_0\Gamma[\beta]} 
\left(\frac{g}{\omega_0}\right)^{\beta-1} e^{- \frac{g}{\omega_0} \beta},
\end{equation}
which matches with the previously known result~\cite{rajesh_m2000} for the uniform case ($\beta=\frac{1}{2}$). 
Similar gamma distribution for the inter-particle spacings has been obtained by Zielen and Schadschneider in 
the context of totally asymmetric RAP with parallel update on an infinite line \cite{zielen_s2002a}. 
As in our case their stationary distribution of the gaps, also depends on jump distribution $R(\eta)$ only through a particular combination of $\mu_1$ and $\mu_2$ given by 
$\lambda_2 = \frac{\mu_1-\mu_2}{\mu_2-\mu_1^2}$, which is analogous to our parameter $\beta$. However, in their case particles are moving on an 
infinite line. As a result their stationary state is completely factorized in contrast to our almost factorized form.

In Fig.\,\ref{fig:p1msrap}, we compare the single site marginal gap/mass distribution in  (\ref{eq:marginalM}) with the same obtained numerically 
for different jump distributions $R(\eta)$. The details are given in the caption. The agreement between theory and numerics is perfect for the power law, 
sine and exponential distributions as expected, since they satisfy condition~\eqref{eq:stat1massltu}.
Finally, for certain classes of jump distribution, such as
$R(\eta)= 6\eta(1-\eta)$ that doesn't satisfy the condition \eqref{eq:stat1massltu},
the fact that the gap distribution still is numerically very close to \eqref{eq:marginalM} (see fig. \ref{fig:p1msrap}: case 'AR') is indeed very interesting. 
Similar (in spirit) results were also found in \cite{rajesh_m_01}.


\subsection{Variance and correlations of the gaps for arbitrary jump distribution $R(\eta)$}
\label{corr-p-eq-q}
\noindent
For arbitrary jump distributions that do not satisfy  (\ref{eq:stat1massltu}), 
the almost factorization of the stationary JPDF for the gaps does not hold in general and finding exact expression of JPDF is somewhat difficult.
However, one can compute the mean, variance and correlations of the gaps in the stationary state for arbitrary $R(\eta)$. Multiplying both sides of the SS master equation 
(\ref{eq:meqss}) by $g_i$ and then integrating over all $g_j$s, one can easily find that the average mass profile $m_i = \langle g_i \rangle$ satisfies 
\begin{equation}
\label{av-ms-p-eq-q}
 m_{i+1}-2m_i+m_{i-1}=(2p-1)(\delta_{i,1}-\delta_{i,N})(m_1-m_N),~~\text{for}~~i=1,2,...,N.
\end{equation}
This equation can be easily solved to get $m_i=\omega_0=L/N$ for all $i$. Similarly multiplying both sides of (\ref{eq:meqss}) by $(g_i-\omega_0)(g_j-\omega_0)$ and 
then integrating over all $g_j$s one obtains the equation satisfied by the correlation function $d_{i,j}=\langle g_i g_j \rangle -\omega_0^2$.
The equation is given by
\begin{eqnarray}
\label{eq:corr-p-eq-q}
         && \mu_1 (4d_{i,j}-d_{i,j+1}-d_{i,j-1}-d_{i+1,j}-d_{i-1,j})= \mu_2
         (\delta_{j,i}-\delta_{j,i+1})~(d_{i+1,i+1}+d_{i,i}) + \mu_2(\delta_{j,i}-\delta_{j,i-1}) (d_{i-1,i-1}+d_{i,i})  \nonumber \\
         && ~~~~~~~~~~~~~~~~~~~~~~~~~~~~~~~~~~~~+\mu_1 (\delta_{i,N}-\delta_{i,1}) (2p-1) (d_{j,1} - d_{j,N}) 
                                                + \mu_1 (\delta_{j,N}-\delta_{j,1}) (2p-1) (d_{1,i} - d_{N,i})  \\
         &&~~~~~~~~~~~~~~~~~~~~~~~~~~~~~~~~~~~~+ \mu_2(\delta_{i,N}-\delta_{i,1})(\delta_{j,N}-\delta_{j,1}) (2 p-1)(d_{N,N}+d_{1,1} ) 
-2~\mu_2~\omega_0^2~(\delta_{j,i+1}-2\delta_{j,i}+\delta_{j,i-1})  \nonumber \\
&&~~~~~~~~~~~~~~~~~~~~~~~~~~~~~~~~~~~~+2 (2 p-1)\omega_0^2 \mu_2(\delta_{i,N}-\delta_{i,1}) (\delta_{j,N}-\delta_{j,1}). \nonumber
\end{eqnarray}
One can try to solve this equation numerically only to observe that $d_{i,j}=D_1\delta_{i,j}+D_2$ where $D_1$ and $D_2$ 
are constants that depend on $\mu_1,~\mu_2$, $\omega_0$ and $N$ but not on $p$. 
Using this form of $d_{i,j}$ in  (\ref{eq:corr-p-eq-q}) one can solve for $D_1$ and $D_2$ to get 
\begin{equation}
 d_{i,j} = \left( \frac{\mu_2 \omega_0^2}{\mu_1+(N-1)(\mu_1-\mu_2)}\right) [N\delta_{i,j}-1],~~~~\text{for}~~~p=q,
\end{equation}
as stated in the introduction. It can be checked that in the case of factorized SS, 
correlations computed from the JPDF \eqref{eq:statbeta1} consistently yields the same expression. 

Using the above knowledge of the fluctuation of the 
gaps one can study the motion of the CM which is defined as $X_{cm}(t)=\frac{1}{N}\sum_{i=1}^N x_i(t)$. From the evolution of $x_i(t)$s in \eqref{dyna-ith} and 
\eqref{dyna-tracer}, one can write the evolution of $X_{cm}(t)$ as $X_{cm}(t+dt)- X_{cm}(t) = \Delta X_{cm}(t) =\frac{1}{N} \sum_{i=1}^{N} \psi_i g_i(t)$, where 
the $\psi_i$s are stationary random variables independent of the $g_i$ and uncorrelated to order $dt$. As a result, the distribution of $X_{cm}(t)$ 
in the large $t$ limit can be computed to be a Gaussian distribution 
with variance $\sim 2 D_{cm} t$ where the diffusion coefficient is given by 
$ D_{cm}=\frac{1}{2 N}~\omega_0^2\frac{\mu_1\mu_2}{\mu_1-\mu_2}$, and this prediction agrees with the 
numerics (not shown here) for large enough values of $p$.

\section{Biased Tracer Case : $p \neq q$}
\label{section:tsrap}
\noindent
In this section we consider $p \neq q$ case where the tracer particle is driven. As a result, in the long 
time limit the system will reach a current carrying steady state which will support an inhomogeneous mass/gap density profile in contrast to the $p=q$ case.
As in the unbiased case here too we are interested in the statistics of the gaps between neighboring particles in the steady state. 
The joint probability distribution of the gaps $P_{N,L}({\bf G})$ in this case will in general not have a factorized form  
as it has in the $p=q$ case for some choices of $R(\eta)$.
Computation of $n$ point joint distributions naturally involve $(n+1)$ point joint distributions, which makes it difficult to determine the JPDF exactly. 
However, we will later see that the correlations among different sites decrease to zero as $\sim 1/N$  in $N \to \infty$ limit. 
Hence, let us proceed with the assumption that the joint probability distribution has the following form
\begin{equation}
\label{MF-JPDF}
 P_{N,L}(g_1,g_2,...,g_N) \simeq  \prod_{i=1}^N P_i(g_i,\omega_0)\left[1+\mathcal{O} \left(\frac{1}{N}\right)\right]~\delta \left(\sum_{i=1}^Ng_i -L \right),
\end{equation}
in the large $L$ and $N$ limit keeping the mass density $\omega_0=L/N$ fixed. Inserting this form of JPDF in  (\ref{eq:meqsrap}) we find 
\begin{eqnarray}
[2+(p+q-1)(\delta_{i,1}+\delta_{i,N})]~P_i (g_i) &=&  \int_0^{\infty} dg_{i-1}' \int_0^{\infty} dg_{i}' \int_0^1
 d\eta~ R(\eta) ~\big{( }[1/2 +(q-1/2)\delta_{i,1}]~\delta[g_i' +\eta g_{i-1}' -g_i] \nonumber \\
&&~~~~~~~~~~~~+ [1/2 +(p-1/2)\delta_{i,1}]~ \delta[g_i' (1-\eta) -g_i] \big{)}~ P_{i-1}(g'_{i-1}) P_i(g_i') \label{bulk-eq-arap} \\
&+& \int_0^{\infty} dg_{i}' \int_0^{\infty} dg_{i+1}' \int_0^1 d\eta~ R(\eta) \big{(} [1/2 +(p-1/2)\delta_{i,N}]~\delta[g_i' +\eta g_{i+1}' -g_i] \nonumber \\
&&~~~~~~~~~~~~+ [1/2 +(q-1/2)\delta_{i,N}]~\delta[g_i' (1-\eta) -g_i] \big{)} P_{i+1}(g'_{i+1}) P_i(g_i'), \nonumber 
\end{eqnarray}
where the explicit $\omega_0$ dependence of $P_i(g_i,\omega_0)$ has been omitted for convenience.
Taking Laplace transform of both sides with respect to $L$ and $g$, we get 
\begin{eqnarray}
2\frac{\tilde{P}_i(s+s')}{\tilde{P}_i(s)} &\approx & \int_0^1d\eta ~R(\eta)~\left[\frac{\tilde{P}_i(s+s'-\eta s')}{\tilde{P}_i(s)} 
+ \frac{\tilde{P}_i(s+s')}{\tilde{P}_i(s)}~\frac{\tilde{P}_i(s+\eta s')}{\tilde{P}_i(s)} \right] + \mathcal{O} (1/N),~~\text{for}~~1< i < N
\label{single-mas-dist-LT-p-neq-q} 
\end{eqnarray}
in the thermodynamic limit, where the notation $\tilde{P}_i(s) = \int_0^{\infty}dg~e^{-sg} P_i(g)$ and 
the approximation : $P_{i \pm 1}(g) \simeq P_i(g) +\mathcal{O}(1/N)$ have been used. In the above equation the argument $s$ in $\tilde{P}_i$ 
appears while taking Laplace transform with respect to $L$ whereas $s'$ appears while taking Laplace transform with respect to $g_i$. 
Note that  (\ref{single-mas-dist-LT-p-neq-q}) is quite similar to  (\ref{eq:stat1masslt}). Hence following the same 
procedure as performed in sec. (\ref{subsection:srapmassstat}) one can find the solution of (\ref{single-mas-dist-LT-p-neq-q}) as 
\begin{equation}
 \tilde{P}_i(s) = (sh_i +1)^{-\beta},~~~\text{where}~~~\beta=\frac{\mu_1-\mu_2}{\mu_2}, \label{p_tilda_s}
\end{equation}
and $h_i$ is site index $i$ dependent constant. Inverting the Laplace transform of $\tilde{P}_i(s)$ we get exactly the same distribution (\ref{eq:marginalMLinf}) 
except $\omega_0$ is now replaced by $h_i\beta$.
This suggests $h_i$ has to be related to the local average gap $\langle g_i \rangle$. Taking first derivative of $\tilde{P}_i(s)$ in (\ref{p_tilda_s}) 
with respect to $s$ and evaluating it at $s=0$,  one can show that $m_i =\langle g_i \rangle = \int dg~g~P_i(g) = h_i\beta$. Hence we have 
\begin{equation}
 P_i(g) = \frac{\beta^{\beta}}{m_i\Gamma[\beta]} \left(\frac{g}{m_i}\right)^{\beta-1} \ee^{- \frac{g}{m_i}\beta}. \label{gap-dist-p-neq-q}
\end{equation}
The question now is: how does the average gap $m_i$ depend on $i$ ? In the next section we explicitly compute this dependence.

\subsection{Average gap/mass profile}
\begin{figure}[!ht]
	\begin{center}
		\includegraphics[width=0.4\textwidth]{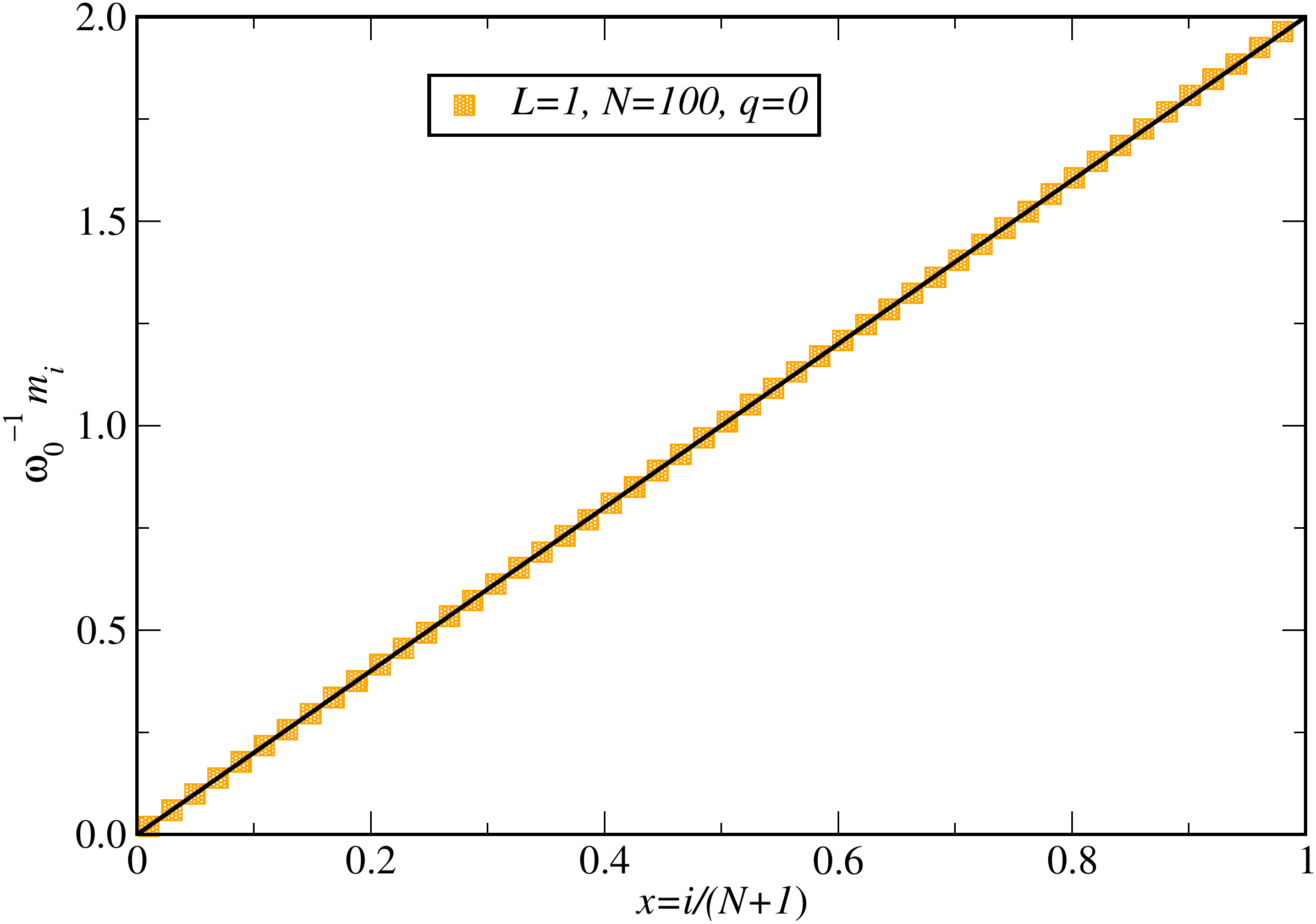} 
	\end{center}
	\caption{\small 
The average stationary mass $m_i$ vs. $x=i/(N+1)$ is plotted for $N=100$ and $L=1$, the squares being numerical and the straight 
solid line is $\mathcal{M}(x)$ in ~\eqref{msqsclng} with $q=0$. In this plot jump distribution is uniform and the first particle is the tracer particle.}  
	\label{fig:mi}
\end{figure}
\begin{figure}[!ht]
	\begin{center}
		\includegraphics[width=0.6\textwidth]{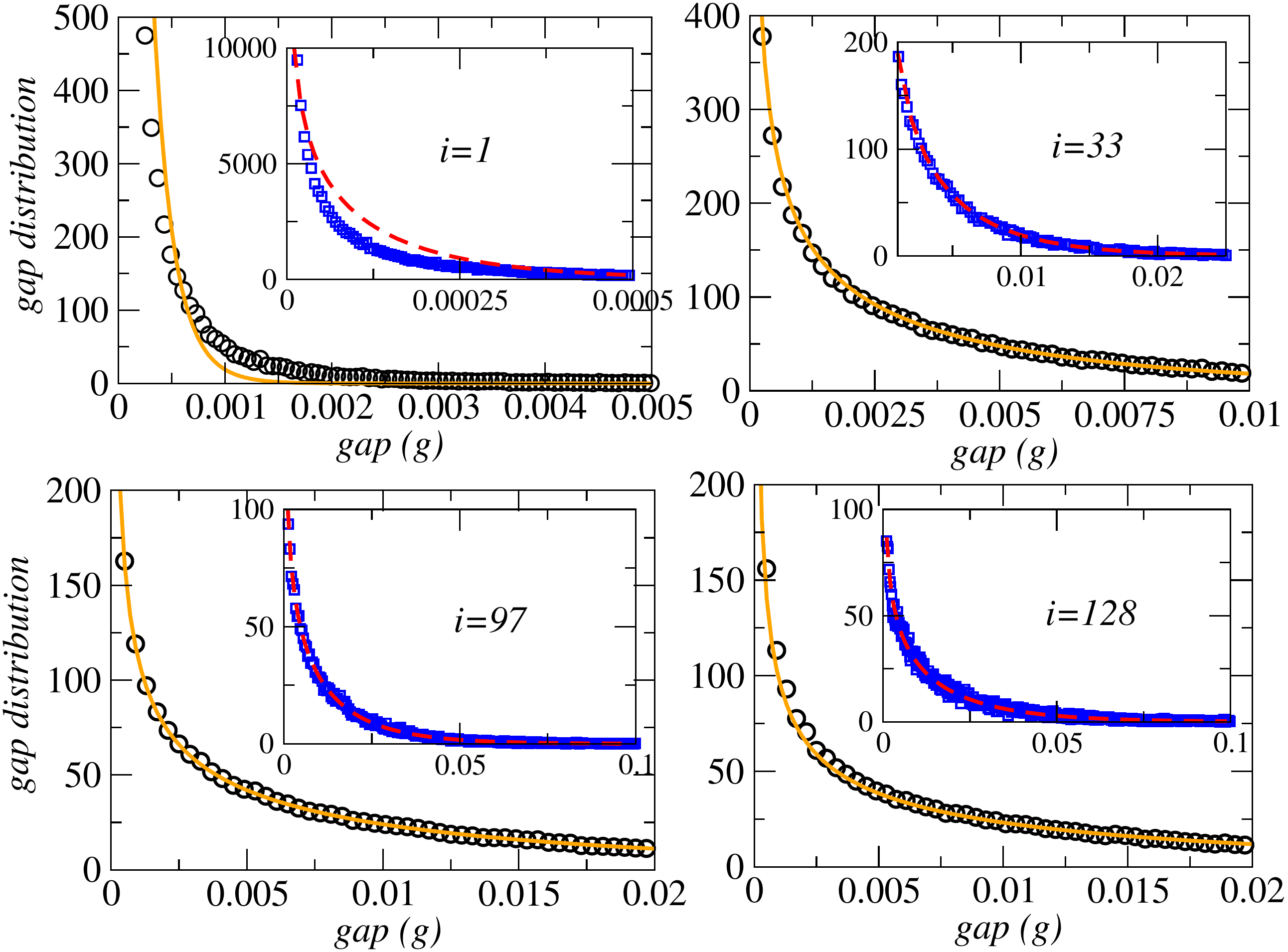} 
	\end{center}
	\caption{\small Single site gap/mass distribution $P^{(1)}_{i}(g)$ for $N=128$, $L=1$ and $i=1,33,97,128$. 
The black circles and the blue squares are numerical measurements for $R(\eta)=1$ and $R(\eta)=6~\eta(1-\eta)$ respectively. The value of $\beta$ corresponding to these two 
distributions are $\beta=1/2$ and $\beta=2/3$. The orange solid and the red dashed lines are the approximate theoretical expressions~\eqref{gap-dist-p-neq-q} 
for $\beta=1/2$ and $\beta=2/3$ respectively, along with $m_i$ from \eqref{eq:istatsrapt}. For all the above plots the jump distribution is uniform and $q=0$. The 
first particle is the tracer particle.} 
	\label{fig:p1mtsrap}
\end{figure}

\noindent
To compute the average mass $m_i=\langle g_i \rangle$ at site $i$, we follow the same procedure as done in the $p=q$ case in sec. (\ref{corr-p-eq-q}).
Multiplying both sides of the stationary state (SS) master equation (\ref{eq:meqss}) by $g_i$ and integrating over all masses $g_j,~j=1,2,\ldots,N$ 
we get the following equation :
\begin{equation}
\label{eq:evomi}
 m_{i+1}-2m_i+m_{i-1}=(\delta_{i,1}-\delta_{i,N})[(2p-1)m_1-(2q-1)m_N),~~\text{for}~~i=1,2,...,N.
\end{equation}
Note that this equation reduces to  (\ref{av-ms-p-eq-q}) for $p=q$.
The solution of the above equation can be easily obtained as 
\begin{equation}
\label{eq:istatsrapt}
m_i= \frac{\omega_0}{(p+q)(N-1)+1}~\left[2(p-q)i+(2qN-2p+1)\right], 
\end{equation}
which (solid line) agrees very nicely with the numerical results (Orange boxes) in Fig.\,\ref{fig:mi}. 
Using this expression of $m_i$, we compare our theoretical prediction \eqref{gap-dist-p-neq-q} 
against the numerical measurements for two choices of $R(\eta)$s in Fig.\,\ref{fig:p1mtsrap}. 
We find quite good match as long as $i$ is not too close to the tracer and at its front. 

\noindent
From the value of the masses the stationary velocity $v^\st$ of the tracer particle can be computed. 
Taking average over $\eta$ variables in \eqref{dyna-tracer} and using \eqref{eq:istatsrapt}, one gets 
\begin{equation}
\label{eq:vstatsrapt}
v^\st = \lim \limits_{dt\to 0} \frac{\langle x_1(t+dt)-x_1(t)\rangle}{dt} 
 = \mu_1 (p~m_{1}-q~m_N) = \omega_0\frac{\mu_1(p-q)}{(p+q)(N-1)+1}.
\end{equation}
Note that the velocity vanishes in the thermodynamic limit as $\sim 1/N$. 
In this limit the expression of $m_i$ takes the following scaling form
\begin{equation}
	m_i = \omega_0~\mathcal{M}\left(\frac{i}{N}\right)~~~\text{where}~~\mathcal{M}(x)= \frac{2}{p+q}~[(p-q)x+q],~~~x\in [0,~1).   \label{msqsclng} 
\end{equation}

\subsection{Fluctuations and correlations of the masses}
\label{fluc-corr}
In section (\ref{section:tsrap}) we have seen that, to the leading order in thermodynamic limit the joint probability distribution 
of the gaps $P_{N,L}({\bf G})$ can be approximated by a factorized form \eqref{MF-JPDF}. 
However, for finite size systems there will be corrections to this factorized form and those corrections arise from the 
correlation among different gaps. In this section we study the correlations of the gaps for $p\neq q$. 
We denote the connected two-point correlation function by $d_{i,j} = \langle g_i g_j \rangle - \langle g_i \rangle \langle g_j\rangle$ which by definition 
is symmetric $d_{i,j} = d_{j,i}$. Also due to the mass conservation property $\sum_{i=1}^{N} g_i(t) = L$, correlation $d_{i,j}$ satisfies 
$\sum_{i=1}^{N} d_{i,j} = 0$. To find the equations satisfied by $d_{i,j}$ in SS, we multiply both sides of equation (\ref{eq:meqss}) by 
$(g_i - \langle g_i \rangle )(g_j - \langle g_j \rangle ) $ and then integrating both sides over all the gaps/masses we get, 
\begin{eqnarray}
         \label{eq:evodij}
         && \mu_1 (4d_{i,j}-d_{i,j+1}-d_{i,j-1}-d_{i+1,j}-d_{i-1,j})= \mu_2
         (\delta_{j,i}-\delta_{j,i+1})~(d_{i+1,i+1}+d_{i,i}) + \mu_2(\delta_{j,i}-\delta_{j,i-1}) (d_{i-1,i-1}+d_{i,i})  \nonumber \\
         &&~~~~~~~~~~~~~~~~~~+\mu_1 (\delta_{i,N}-\delta_{i,1}) (d_{j,1}(2p-1)  - (2q-1) d_{j,N}) + \mu_1 (\delta_{j,N}-\delta_{j,1}) (d_{1,i} (2p-1) - (2q-1) d_{N,i}) \\
         &&~~~~~~~~~~~~~~~~~~+ \mu_2(\delta_{i,N}-\delta_{i,1})(\delta_{j,N}-\delta_{j,1}) (d_{N,N}(2 q-1)+(2p-1) d_{1,1} ) 
                             + \mu_2(\delta_{j,i}-\delta_{j,i+1}) (m_{i+1}^2 + m_i^2)  \nonumber \\
         &&~~~~~~~~~~~~~~~~~~+ \mu_2(\delta_{j,i}-\delta_{j,i-1})  (m_{i-1}^2 + m_i^2)
                             + \mu_2(\delta_{i,N}-\delta_{i,1}) (\delta_{j,N}-\delta_{j,1}) (m_{N}^2(2 q-1) + (2 p-1) m_1^2), \nonumber
 \end{eqnarray}
where $m_i$ is given in  (\ref{eq:istatsrapt}). One now needs to solve this equation for $d_{i,j}$. 
\begin{figure}[!ht]
	\begin{center}
		\includegraphics[width=0.4\textwidth]{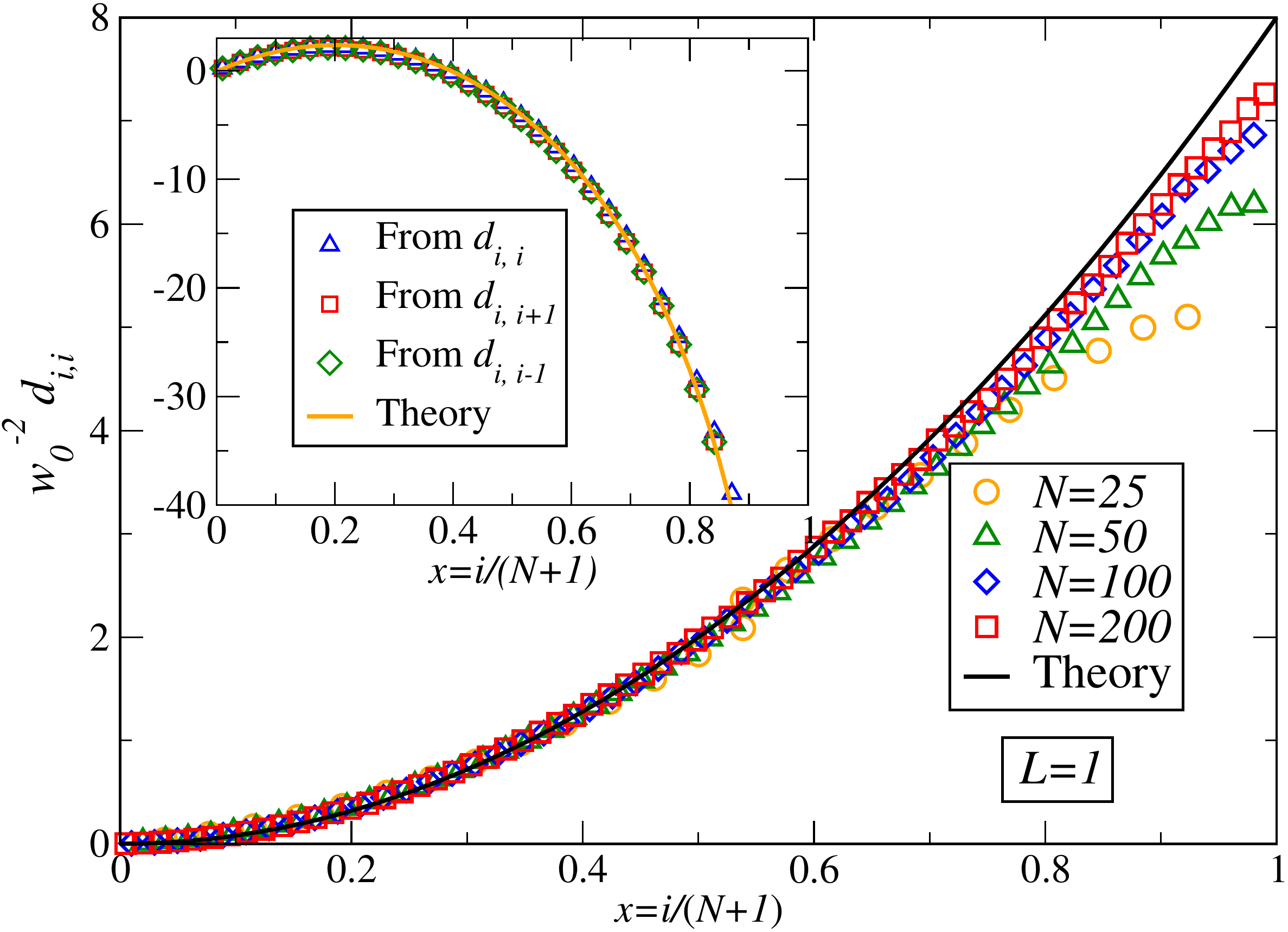} 
	\end{center}
	\caption{\small Rescaled stationary mass variance profile $\omega_0^{-2} d_{i,i}$ as a function 
of the rescaled coordinate $x=\frac{i}{N+1}$ for $N=50$, $100$, $200$ and $L=1$. The circles are 
numerical results with $q=0$ and the black solid line is the corresponding theoretical 
expression~\eqref{eq:C0}. In the inset we verify the relation \eqref{eq:C1} numerically. The blue triangles correspond 
$N~\omega_0^{-2}~[d_{i,i}-\omega_0^2\mathcal{C}_0(i/N)]$ vs.$x$, the red squares correspond 
$N~\omega_0^{-2}~[\mu_1/(\mu_1-\mu_2)]~d_{i,i+1}$ vs. $x$ and the green diamonds correspond to $N~\omega_0^{-2}~[\mu_1/(\mu_1-\mu_2)]~d_{i,i-1}$ vs. $x$. 
The solid orange line corresponds to the theoretical $[\mu_1/(\mu_1-\mu_2)]~\mathcal{D}(x,x)$. 
For both the plots, the first particle is the tracer particle and the jump distribution  is uniform : $R(\eta)=1$.}  
	\label{fig:di}
\end{figure}
\begin{figure}[!ht]
	\begin{center}
		\includegraphics[width=0.6\textwidth]{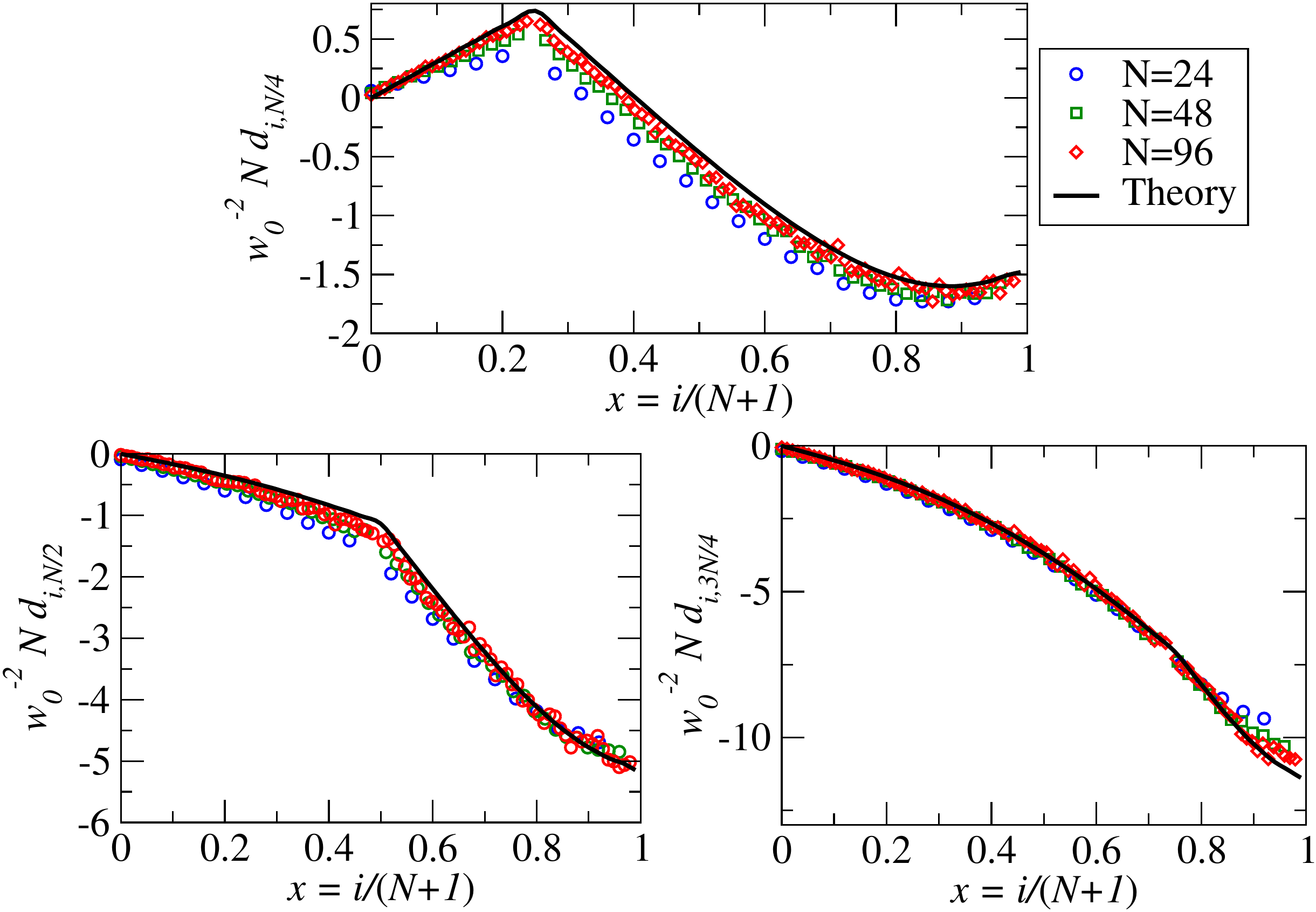} 
	\end{center}
	\caption{\small Rescaled stationary non-diagonal mass-mass correlation $\omega_0^{-2}~N~d_{i,j}$ as a function 
of the rescaled coordinate $x=\frac{i}{N+1}$ for $j=N/4,~N/2$ and $3N/4$ with $N=24$, $48$, $96$ and $L=1$. 
The circles are numerically obtained and the black solid line is the theoretical 
expression~~\eqref{sol-Dxy}. The first particle is the tracer particle which has $q=0$. 
Jump distribution is uniform $R(\eta)=1$. }  
	\label{fig:nondi}
\end{figure}

\noindent
Before discussing the solution of equation \eqref{eq:evodij}, let us present our simulation results. 
Numerical measurements of the two-point correlation function are plotted in Figs. (\ref{fig:di}) and (\ref{fig:nondi}). In Fig. (\ref{fig:di}) we 
plot the scaled stationary mass variance $\omega_0^{-2} d_{i,i}$ as function 
of the rescaled coordinate $\frac{i}{N+1}$ for $N=50$, $100$, $200$ and $L=1$, whereas in fig. (\ref{fig:nondi}) we plot 
scaled stationary non-diagonal mass-mass correlation $\omega_0^{-2}~N~d_{i,j}$ as a function 
of the rescaled coordinate $\frac{i}{N+1}$ for $j=N/4,~N/2$ and $3N/4$ with $N=24$, $48$, $96$ and $L=1$.
In both figures we find nice data collapse, which implies that the diagonal correlations \emph{i.e.} fluctuations $d_{i,i}$ 
and the non-diagonal correlations scale differently in the thermodynamic limit. 
In fact the numerical results indicate the following scaling behaviors
\begin{eqnarray}
 d_{i,j}&=& \frac{\omega_0^2}{N}~\mathcal{D}\left(\frac{i}{N},~ \frac{j}{N}\right) + o(N^{-1})
 \qquad\text{for}~~i\neq j,\nonumber\\
&&  \label{scaling-corr} \\
 d_{i,i} &=& \omega_0^2~\mathcal{C}_0\left(\frac{i}{N} \right) +  \frac{\omega_0^2}{N}~\mathcal{C}_1\left(\frac{i}{N}\right) 
+\frac{\omega_0^2}{N}~\mathcal{D}\left(\frac{i}{N},~ \frac{i}{N}\right)  + o(N^{-1}),
\nonumber
\end{eqnarray}
where $o(\ell)$ represents terms of orders smaller than $\ell$. We thus look for solution of  (\ref{eq:evodij}) in the forms (\ref{scaling-corr}). 
We insert the scaling form of $m_i$ from ~\eqref{msqsclng} and of $d_{i,j}$ from ~\eqref{scaling-corr} in (\ref{eq:evodij}).
Expanding both sides of (\ref{eq:evodij}) as powers of $1/N$ keeping $\omega_0=L/N$ fixed and equating terms of same power, we find that  
order $N^{-1}$ and $N^{-2}$ give
\begin{eqnarray}
\mathcal{C}_0(x)&=&\frac{\mu_2}{\mu_1-\mu_2}~\mathcal{M}^2(x) = \frac{\mu_2}{\mu_1-\mu_2}~\frac{4}{(p+q)^2}~[(p-q)x+q]^2, \label{eq:C0}
\end{eqnarray}
and whereas order $N^{-3}$ yields the following differential equation:
\begin{eqnarray}
&&\left( \partial_x^2 + \partial_y^2  \right) \mathcal{D}(x,y)
 = -\frac{\mu_2}{\mu_1-\mu_2}~\delta(x-y)~\frac{d^2}{dx^2}\mathcal{M}(x)^2 
= -\frac{8\mu_2}{\mu_1-\mu_2}~\left(\frac{p-q}{p+q}\right)^2~\delta(x-y)[1-\delta(x-1)], \label{sclng-diff-eq} \\
&&\text{with~boundary~conditions,} \nonumber \\
&&~~~~~~~~~~~~~~~~~~~~~~~~~~~~~~~~~~~~~~~~~~~~~~~~~p~\mathcal{D}(x,y)|_{x \to 0} =q~\mathcal{D}(x,y)|_{x \to 1}, \nonumber \\
&&~~~~~~~~~~~~~~~~~~~~~~~~~~~~~~~~~~~~~~~~~~~~~~~~~p~\mathcal{D}(x,y)|_{y \to 0} =q~\mathcal{D}(x,y)|_{y \to 1}, \nonumber \\
&&~~~~~~~~~~~~~~~~~~~~~~~~~~~~~~~~~~~~~~~~~~~~~~~~~\partial_x \mathcal{D}(x,y)|_{x \to 0} =\partial_x \mathcal{D}(x,y)|_{x \to 1},\label{Dxy-bcs} \\
&&~~~~~~~~~~~~~~~~~~~~~~~~~~~~~~~~~~~~~~~~~~~~~~~~~\partial_y \mathcal{D}(x,y)|_{y \to 0} =\partial_y \mathcal{D}(x,y)|_{y \to 1}. \nonumber
\end{eqnarray} 
The details of the derivation of the above equations are given in appendix (\ref{derivation-sclng-diff-eq}). 
Equations~\eqref{sclng-diff-eq} and~\eqref{Dxy-bcs} have to be supplemented by the vanishing integral condition
\begin{equation}
\label{eq:Dxyint0}
\mathcal{C}_0(x) + \int_{y=0}^1 \mathcal{D}(x,y) d y = 0,
\end{equation} 
which comes directly from the fact that the sum of $d_{i,j}$ on a full row vanishes due to mass conservation.

Note that the leading term $\mathcal{C}_0(x)=\frac{4 \mu_2}{\mu_1-\mu_2}~\mathcal{M}(x)^2$ 
for the diagonal correlation $d_{i,i}$ in  (\ref{scaling-corr}) can also be obtained directly from the distribution (\ref{gap-dist-p-neq-q}). 
It can also be obtained by writing the discrete equation (\ref{eq:evodij}) for $i=j$ and neglecting any non-diagonal correlation (as they are 
order $\mathcal{O}(1/N)$ smaller than the diagonal correlation ) in thermodynamic limit. This way of obtaining the continuum description 
also provides information about the terms at the next order ($\sim 1/N$) in the second line of \eqref{scaling-corr}, (see appendix \ref{derivation-sclng-diff-eq})
\begin{equation}
 \mathcal{C}_1(x)=\frac{ \mu_2}{\mu_1-\mu_2}~\mathcal{D}(x,x), \label{eq:C1} 
\end{equation}
In fig. (\ref{fig:di}) we compare theoretical solution for $d_{i,i}$ at leading order \emph{i.e.} $\mathcal{C}_0(x)$ from \eqref{eq:C0},  
with the numerical results (circles) and observe very nice agreement. In the inset of the same figure we verify \eqref{eq:C1} numerically.

We now turn our attention back to solve the differential equation (\ref{sclng-diff-eq}) satisfied by the scaling function $\mathcal{D}(x,y)$ associated to non-diagonal 
correlation. Equation (\ref{sclng-diff-eq}) can be interpreted as Poisson equation where $\mathcal{D}(x,y)$ is the electric potential created inside the square domain
 $(x,y) \in [0,1]^2$ by the charge distribution along the diagonal. Similar interpretation have been used in \cite{Sha_m_m2014} in the context of long range 
density-density correlation in diffusive lattice gas with simple exclusion interaction. 
The boundary conditions~~\eqref{Dxy-bcs} on the derivatives being the same at $0$ and $1$ imply that the total flux of the electric field exiting 
the square domain vanishes, so that the charge inside the square must vanish as well. One can directly observe this fact by integrating the charge distribution on 
the right hand side of \eqref{sclng-diff-eq} over the unit square domain.

In general, to solve an inhomogeneous differential equations one usually defines a Green's function as the solution of 
$\nabla^2~\mathcal{G}(x,y|x_0,y_0)= \delta(x-x_0)\delta(y-y_0)$ with boundary conditions same as the of the original inhomogeneous differential equation. 
However we cannot use this Green's function to compute $\mathcal{D}(x,y)$ because this corresponds to potential created by a net unit charge. 
We therefore need to compensate the $\delta(x-x_0)\delta(y-y_0)$ charge by a negative unit charge distributed on the square. 
We choose the following charge distribution inside the square  $(x,y) \in [0,1]^2$
\begin{equation}\label{eq:Gf}
 \nabla^2~\mathcal{G}_{q/p}(x,y|x_0,y_0)= \delta(x-x_0)\delta(y-y_0) -\frac{4}{(p+q)^2} (p x + q (1-x))(p y + q (1-y)), 
\end{equation}
with boundary conditions~(\ref{Dxy-bcs}). The purpose of choosing this particular charge distribution will become clear in appendix~(\ref{derivation-Green-func}), where  
we provide the details of the solution $\mathcal{G}_{q/p}(x,y|x_0,y_0)$ of the above equation for arbitrary $p$ and $q$.  
For simplicity, here we restrict ourselves to the $q=0$ case for which the Green's function reads  
\begin{eqnarray}
&&\mathcal{G}_0(x,y|x_0,y_0) = 2x\sum_{m=1}^{\infty} \left[ \frac{\sin(2 \pi m y)\cos(2 \pi m y_0)}{\pi^3m^3} 
- \frac{y\cos(2 \pi m y)\cos(2 \pi m y_0)+(1-y_0)\sin(2 \pi m y)\sin(2 \pi m y_0)}{\pi^2m^2}\right]  \nonumber \\
&&~~~~~~~~~~~~~~~~~~~~+4\sum_{m=1}^{\infty} F_{2\pi m}(y|y_0)~\left[~x\cos(2 \pi m x)\cos(2 \pi m x_0) + (1-x_0)\sin(2 \pi m x)\sin(2 \pi m x_0)~\right] \label{Green-func} \\
&&~~~~~~~~~~~~~~~~~~~~+16\pi \sum_{m=1}^{\infty} m~\sin(2 \pi m x)\cos(2 \pi m x_0)~\Sigma_{2 \pi m}(y|y_0), \nonumber \\
&& \text{where} ~~~~~~~~~~~~~~~~
F_a(y|y_0)=\frac{1}{a(1-\text{cosh}[a])}
\begin{cases}
&\text{cosh}[a(1-y_0)]~  \text{sinh}[ay],~~~~~~~~~~~~~~~~~~~~~~~~~~~~~~~y\leq y_0 \\            
&\text{sinh}[a(y-y_0)]+\text{sinh}[ay_0]~\text{cosh}[a(1-y)],~~~~~~~~~y>y_0
\end{cases},\label{F_a} \\ 
&&~~~~~~~~~~~~~~~~~~~~~~~~\text{and}~~~~~
\Sigma_a(y|y_0)=\int_0^1dz~F_a(y|z)F_a(z|y). \label{Sigma}
\end{eqnarray}
Once we know the Green function $\mathcal{G}_0(x,y|x_0,y_0)$, the general solution $\mathcal{D}(x,y)$ is given by the sum of a particular solution expressed in 
terms of the Green's function 
and a homogeneous solution of the equation $V_0 x y$, where $V_0$ is an integration constant determined using the integral condition~~\eqref{eq:Dxyint0}.
The complete solution of \eqref{sclng-diff-eq}-\eqref{Dxy-bcs}-\eqref{eq:Dxyint0} for $q=0$ is finally given by  
\begin{equation}
\mathcal{D}(x,y)= \frac{16\mu_2}{3(\mu_1-\mu_2)}~xy~-~\frac{8\mu_2}{\mu_1-\mu_2}~\int_0^1dx_0\int_0^1dy_0~\mathcal{G}_0(x,y|x_0,y_0)~\delta(x_0-y_0))~[1-\delta(x_0-1)],
\label{sol-Dxy}
\end{equation}
where the first term is the homogeneous part and the second term is the particular solution. 
In Fig.\,\ref{fig:nondi} we compare the above theoretical solution (solid black line) for $\mathcal{D}(x,y)$  along with the Green's function in  (\ref{Green-func})  
with the numerical results (circles). Nice agreement between the two provides verification of our analytical result.

Once again using the knowledge of the gap fluctuations one could intend to study the distribution of the center of mass position $X_{cm}(t)$ as done in the unbiased case.
However the predicted diffusion coefficient does not match with the numerical simulation results (not presented here). 
This discrepancy can be attributed to strong correlations among increments $\Delta X_{cm}(t)$ at different times which can be evidenced numerically.

\section{Conclusion}
\label{section:ccl}

In this paper we have studied the statistics of the gaps between neighboring particles moving according to RAP on a ring. 
The particles hop symmetrically on both sides except 
for a tracer particle which may be driven. Taking advantage of the mapping between the RAP and a MT model, equivalent up to a global shift, we have obtained some 
exact results. In this mapping the masses correspond to the gaps between successive particles of the RAP, and a global constraint therefore enforces that 
the sum of the masses is constant and is equal to the length of the ring.

In the non-driven tracer case ($p=q$) we found that  when the jump distribution $R(\eta)$ satisfies a necessary and sufficient condition (\ref{eq:stat1massltu}) 
the stationary joint distributions of the gaps takes an universal form (\ref{eq:statbeta1}) parametrized by a positive constant $\beta=\frac{\mu_1-\mu_2}{\mu_2}$. 
We showed that there exists infinite number of solutions of (\ref{eq:stat1massltu}) for $R(\eta)$. 
We have found a quite large number of solutions however those do not include all the possible solutions. 
The calculation and qualitative results are very similar to those obatained by Coppersmith \emph{et~al} \cite{coppersmith1996} 
in the context of the q-model of force fluctuations in granular media and to those obtained by Zielen and Schadschneider~\cite{zielen_s2002a} 
for the totally asymmetric RAP with parallel update on an infinite line. In their case also the factorized stationary distributions 
depend only on one parameter, which is however different from ours. Although these distributions are 
never equilibrium ones, cancellations in the master equation still occur between nearest neighbors only and there 
is no large-scale mass current. From this joint stationary distribution the single site marginal distribution of gaps/masses have been computed, 
which also has universal form. 
Interestingly, we numerically observe that, for some $R(\eta)$s which do not satisfy (\ref{eq:stat1massltu}) the single site mass/gap distribution 
can also be described quite well by this universal form.

In the case with a driven tracer the stationary distribution of the masses is not in the form of factorized joint distribution.
In the thermodynamic limit progress has however been made using the fact that the correlations between different masses become negligible 
for large systems. Besides the average 
mass profile, which is easily obtained for any size, we computed the two-point gap correlations in the thermodynamic limit. In the same limit we also have computed 
approximate single site mass distribution of the gaps.

There are several other future extensions of our work. For example, we have already mentioned the problem of characterizing all the solutions $R(\eta)$s of 
(\ref{eq:stat1massltu}) properly. For those jump distributions $R(\eta)$ that do not satisfy (\ref{eq:stat1massltu}), it would be interesting to look for other 
exact forms of the stationary state such as matrix products~\cite{zielen_s2003} or 
cluster-factorized states~\cite{Cha_P_M2015}. 
Similarly, computing the distribution of $X_{cm}(t)$ for the driven tracer case remains an interesting future problem, 
and solving this problem probably requires studying the whole history of 
the system.

A different possibility is to vary the nature of the drive. Note that the $p=q=0$ case of this paper corresponds to a closed segment with fixed walls at $x=0$ and $x=L$,
For this case a large class of fraction distributions $R(\eta)$ have been presented for which the stationary state factorizes. 
It would be interesting to design natural open boundaries for the RAP in order to study, 
say, its conduction properties, and in particular the influence of $\beta$.
Another possible extension would be to introduce a second driven tracer. In this case an important question is to find the nature and the magnitude of the interaction 
between the two tracers.

\section*{Acknowledgments}

We thank Shiladitya Sengupta for carefully reading the manuscript. The support of the Israel Science Foundation (ISF) and of the Minerva
Foundation with funding from the Federal German Ministry for Education and Research is gratefully acknowledged. S.N.M wants to thank the hospitality
of the Weizmann Institute where this work started during his visit as a Weston visiting professor.

\appendix
\section{Some examples of $\hat{\phi}(s)$ in the rational form}
\label{appendix-phihat_s}
Here we present some solutions of  (\ref{eq:stat1masslty}) in $\hat{\phi}(s)=\frac{H_u(s)}{H_b(s)}$ form for large $\beta$:
\begin{eqnarray}
\hat{\phi}(s)&=&\frac{2+es}{1+es+s^2},~~e\geq 2, \nonumber \\
\hat{\phi}(s)&=&\frac{s \sqrt{a^2-4 b+4}-3 a s-2 (b-1) s^2-4}{s^3 \sqrt{a^2-4 b+4}-a \left(s^2+2\right) s-2 b s^2-2},~~a \leq 4,~ b+3< 2a,
~~\text{and}~~ a>4,~4b \leq (4+a^2).
\end{eqnarray}

\section{Derivation of  (\ref{sclng-diff-eq})}
\label{derivation-sclng-diff-eq}
\noindent
To derive the differential equation (\ref{sclng-diff-eq}), 
we first assume the following expansions of the correlation in $1/N$ for large $L$ and $N$ keeping $L/N=\omega_0$ fixed :
{\small
\begin{eqnarray}
\label{corr-sclng}
&&d_{i,j}= \omega_0^2~\left[\frac{1}{N}~\mathcal{D}\left(\frac{i}{N},~ \frac{j}{N}\right) 
+ \frac{1}{N^2}~\bar{\mathcal{D}}\left(\frac{i}{N},~ \frac{j}{N}\right) + o(N^{-2}) \right] \label{sclng-dij}\\
&&d_{i,i} = \omega_0^2~\Big[ \mathcal{C}_0\left(\frac{i}{N} \right) + \frac{1}{N}~\left\{ \mathcal{C}_1\left(\frac{i}{N}\right) 
+ \mathcal{D}\left(\frac{i}{N},~ \frac{i}{N}\right)\right\} \label{sclng-dii} \\
&&~~~~~~~~~~~~~~~~~~~~~~~~~~~~~~~+\frac{1}{N^2}~\left\{ \mathcal{C}_2\left(\frac{i}{N}\right) 
+ \bar{\mathcal{D}}\left(\frac{i}{N},~ \frac{i}{N}\right)\right\} + o(N^{-2})\Big], \nonumber
\end{eqnarray}
}
\noindent
where $\mathcal{C}_1(x)$, $\mathcal{C}_2(x)$ and  $\bar{\mathcal{D}}(x,y)$ are scaling functions at sub-leading order

In the second step, focusing only on the bulk equations (\emph{i.e.} leaving the boundary equations ) we write  (\ref{eq:evodij}) in the 
following discrete Laplace equation form :
\begin{eqnarray}
&&(\Delta d)_{i,j}=\delta_{i,j} A(\{d_{i,j}\})+\delta_{i+1,j}B_1(\{d_{i,j}\})+\delta_{i+1,j}B_{-1}(\{d_{i,j}\}),~~\text{where}, \label{discrete-Lap-eq} \\
&&(\Delta d)_{i,j}=\mu_1 (d_{i,j+1}+d_{i,j-1}+d_{i+1,j}+d_{i-1,j}-4d_{i,j}),\label{discrete-Lap} \\
&&A(\{d_{i,j}\})=-\mu_2(d_{i+1,i+1}+2d_{i,i}+d_{i-1,i-1})-\mu_2(m_{i+1}^2+2m_{i}^2+m_{i-1}^2) \nonumber \\
&&B_1(\{d_{i,j}\})= \mu_2(d_{i,i}+d_{i+1,i+1})+\mu_2(m_{i}^2+m_{i+1}^2) \nonumber \\
&&B_{-1}(\{d_{i,j}\})= \mu_2(d_{i,i}+d_{i-1,i-1})+\mu_2(m_{i}^2+m_{i-1}^2) \nonumber
\end{eqnarray}

We now put the scaling forms (\ref{sclng-dij}) and (\ref{sclng-dii}) of $d_{i,j}$, and the scaling form (\ref{msqsclng}) of $m_i$ 
in the above equation. Equating the terms at order $N^{-1}$, one obtains
\begin{eqnarray}
&&\mathcal{C}_0(x)= \frac{\mu_2}{\mu_1-\mu_2}~\mathcal{M}^2(x),\label{expr-C_01}
\end{eqnarray}
which has been verified numerically in fig. (\ref{fig:di}). After imposing~\eqref{expr-C_01} we can check that terms of order $N^{-2}$ automatically vanish.
The order $N^{-3}$ of (\ref{discrete-Lap-eq}) gives a Laplacian of $\mathcal{D}$, while diagonal terms on the left and right hand sides combine to give a source term,
\begin{equation}
 (\partial_x^2+\partial_y^2)\mathcal{D}(x,y) = - \frac{\mu_2}{\mu_1}\delta(x-y)~[(\mathcal{M}^2)''(x)+\mathcal{C}_0''(x)]
=-\frac{\mu_2}{\mu_1-\mu_2}~\delta(x-y)~(\mathcal{M}^2)''(x),
\label{eq:diff-corr}
\end{equation}
where $f'$ is the derivative of the function $f$. While taking the double derivative of $\mathcal{M}(x)$ one should keep in mind 
that the function changes discontinuously when going from $x=1^-$ (which is actually $x=0^-$ on a ring) to $x=0^+$.  This discontinuity in 
$\mathcal{M}(x)$ can be easily visualized if we consider middle particle ($\frac{N}{2}^{\text{th}}$ particle) as the driven tracer particle 
instead of the $1^{\text{st}}$ particle in RAP. In this case the discontinuity in the average gap/mass profile is at the middle point. 
For example with $q=0$ and $p=\frac{1}{2}$ one can show that  
\begin{equation}
m_i = \frac{2 L}{N(N+1)} \begin{cases}
                              & i+1+\frac{N}{2},~~1\leq i<\frac{N}{2} \\
                              & i+1-\frac{N}{2},~~\frac{N}{2} \leq i \leq N,
                             \end{cases}
\end{equation}
which in the continuum limit would imply $\mathcal{M}(x)=2x - \text{sgn}(x-1/2)$. Hence for $q=0$ the source term of the differential equation (\ref{eq:diff-corr}) is 
given by $-\frac{8\mu_2}{\mu_1-\mu_2}~~\delta(x-y)~[1-\delta(x-1)]$ where other terms that may appear while taking derivative of $\mathcal{M}(x)$ have been omitted as they 
do not contribute to $\mathcal{D}(x,y)$. Similarly for arbitrary $p$ and $q$ one can show 
\begin{equation}
 (\partial_x^2+\partial_y^2)\mathcal{D}(x,y) =-\frac{8\mu_2}{\mu_1-\mu_2}~\left(\frac{p-q}{p+q} \right)^2~\delta(x-y)~[1-\delta(x-1)],
\label{eq:diff-corr-1}
\end{equation}
To obtain the boundary conditions of (\ref{eq:diff-corr-1}), let us now look at the following boundary equations
\begin{eqnarray}
 &&(2p+3)d_{1,j}=d_{2,j}+d_{2,j-1}+d_{N,j+1}+2q~d_{N,j}~,~~3\leq j \leq N-1, \label{d1j}\\
&& (2q+3)d_{N,j}=d_{N,j-1}+d_{N,j+1}+d_{N-1,j}+2p~d_{1,j}~,~~2 \leq j \leq N-2.\label{dNj}
\end{eqnarray}
Again putting the scaling form in  (\ref{sclng-dij}) and equating terms at orders $N^{-k}$ for $k=1$ and $2$ to zero, one gets 
\begin{eqnarray}
p~\mathcal{D}(x,y)|_{x \to 0} &=&q~\mathcal{D}(x,y)|_{x \to 1}, \nonumber \\
p~\mathcal{D}(x,y)|_{y \to 0} &=&q~\mathcal{D}(x,y)|_{y \to 1}, \label{BCs} \\
\partial_x \mathcal{D}(x,y)|_{x \to 0} &=&\partial_x \mathcal{D}(x,y)|_{x \to 1} 
, \nonumber  \\
\text{and},~~~\partial_y \mathcal{D}(x,y)|_{y \to 0} &=&\partial_y \mathcal{D}(x,y)|_{y \to 1} 
. \nonumber
\end{eqnarray}
While obtaining the above relations we look at terms until order $N^{-2}$ because 
had we included these boundary equations in (\ref{discrete-Lap-eq}) they 
would appear with Kronecker deltas $\delta_{i,1}$ or $\delta_{i,N}$ which would provide an extra $\frac{1}{N}$ in the continuum limit.
Similarly one can check that, given  (\ref{expr-C_01}), the remaining equations (for some specific points at boundaries) are automatically satisfied up to order $N^{-3}$.

To prove  (\ref{eq:C1}) for $\mathcal{C}_1(x)$ which provides the next order term for $d_{i,i}$ in (\ref{sclng-dii}), we start with the discrete equation for 
$d_{i,i}$ from (\ref{eq:evodij}). Putting the scaling forms for $m_i$ and $d_{i,j}$ and using $\mathcal{C}_0(x)= \frac{\mu_2}{\mu_1-\mu_2}~\mathcal{M}^2(x)$ one finds
\begin{equation}
 \mathcal{C}_1(x)=\frac{\mu_2}{\mu_1-\mu_2}~\mathcal{D}(x,x)
\end{equation}
 at order $\mathcal{O}(1/N)$. In the inset of fig. (\ref{fig:di}) we provide numerical verification of this relation.

\section{Derivation of the Green function $\mathcal{G}_{q/p}(x,y|x_0,y_0)$ in  (\ref{Green-func})}
\label{derivation-Green-func}

In this appendix we compute the Green's function that appears in the determination of the off-diagonal correlations~~\eqref{Green-func}. 
The Laplacian with boundary conditions \eqref{BCs}, is not a self-adjoint operator. Hence the solution can not be expanded in the usual $\sin x$ and $\cos x$ basis.
Here we extend the method used by Bodineau \textit{et al.}~(\cite{bodineau_d_l2010}, appendix A) to our two-dimensional case and consider the following 
basis functions 
\begin{eqnarray}
f_k^{(1)}(x) &=& 2(p x +q (1-x)) \cos(2 \pi k x), \qquad f_k^{(2)}(x) =\sin(2 \pi k x) \\
g_k^{(1)}(x_0) &=& \cos(2 \pi k x_0), \qquad g_k^{(2)}(x_0) = 2(p(1-x_0)+q x_0)\sin(2 \pi k x_0). \nonumber
\end{eqnarray}
The functions $f_k^{(1)}(x)$ and $f_k^{(2)}(x)$ are independent, each of them satisfies the boundary conditions~~\eqref{BCs} at fixed $y$, 
and their second derivatives can be expressed as 
\begin{eqnarray}
(f_k^{(1)})'' &=& -(2 \pi k)^2 f_k^{(1)} - 8 (p-q) \pi k f_k^{(2)}, \nonumber \\
(f_k^{(2)})'' &=& -(2 \pi k)^2 f_k^{(2)}.
\end{eqnarray}
 In terms of these functions one can check that for $(x,x_0)\in[0,1]^2$ delta function $\delta(x-x_0)$ can be expanded as 
\begin{equation}
\label{eq:expdelta}
\delta(x-x_0) = \frac{2}{p+q} \sum_{k=0}^\infty \left[ a_k^{(1)} f_k^{(1)} (x) g_k^{(1)} (x_0) + a_k^{(2)} f_k^{(2)} (x) g_k^{(2)} (x_0)\right],
\end{equation} 
where
\begin{eqnarray}
a_k^{(1)} &=& 1-\frac{\delta_{k,0}}{2}, \qquad a_k^{(2)} = 1. \label{a_k} 
\end{eqnarray}
We may therefore look for a solution of the form
\begin{equation}
\label{eq:ansatzG}
\mathcal{G}_{q/p}(x,y|x_0,y_0) = \frac{4}{(p+q)^2} \sum_{(k,l)\not \equiv (0,0)}^\infty 
\sum_{i,j=1}^2\left[ c_{k,l}^{i,j} a_k^{(i)} f_k^{(i)} (x) g_k^{(i)} (x_0) a_l^{(j)} f_l^{(j)} (y) g_l^{(j)} (y_0)\right],
\end{equation}
where the $c_{k,l}^{i,j}$ are the unknowns to be determined. Note that in the above expansion the $(0,0)$ term is not there. 
Since the contribution of such term on LHS is zero, we have to avoid the presence of such term on the RHS too. That is done by conveniently adding  
$-\frac{4}{(p+q)^2}~[p x + q (1-x)]~[p y + q (1-y)]$ to the delta source in \eqref{eq:Gf}.
Putting the form~~\eqref{eq:ansatzG} in the equation for the Green 
function~~\eqref{eq:Gf}, we obtain an algebraic equation for each value of $(k,l) \not \equiv (0,0)$ which after solving provide 
\begin{eqnarray}
\label{coeff-GF}
c_{k,l}^{1,1} &=& -\frac{1}{4 \pi^2 (k^2+l^2)}, \nonumber \\
c_{k,l}^{1,2} &=& -\frac{1}{4 \pi^2 (k^2+l^2)}  + 2 (p-q) \frac{a_l^{(1)} g_l^{(1)}(y_0)}{a_l^{(2)} g_l^{(2)}(y_0)} \frac{l}{4 \pi (k^2+l^2)^2}, \nonumber \\
c_{k,l}^{2,1} &=& -\frac{1}{4 \pi^2 (k^2+l^2)}  + 2 (p-q) \frac{a_k^{(1)} g_k^{(1)}(x_0)}{a_k^{(2)} g_k^{(2)}(x_0)} \frac{k}{4 \pi (k^2+l^2)^2}, \\
c_{k,l}^{2,2} &=& -\frac{1}{4 \pi^2 (k^2+l^2)}  + 2 (p-q) \frac{a_k^{(1)} g_k^{(1)}(x_0)}{a_k^{(2)} g_k^{(2)}(x_0)} \frac{k}{4 \pi (k^2+l^2)^2} \nonumber \\
&&+ 2 (p-q) \frac{a_l^{(1)} g_l^{(1)}(y_0)}{a_l^{(2)} g_l^{(2)}(y_0)} \frac{l}{4 \pi (k^2+l^2)^2} 
- 8 (p-q)^2 \frac{a_l^{(1)} g_l^{(1)}(y_0)}{a_l^{(2)} g_l^{(2)}(y_0)} \frac{a_k^{(1)} g_k^{(1)}(x_0)}{a_k^{(2)} g_k^{(2)}(x_0)} \frac{k l}{4 \pi (k^2+l^2)^3}. \nonumber
\end{eqnarray} 

Note that Green function $\mathcal{G}_{q/p}(x,y|x_0,y_0)$ in \eqref{eq:ansatzG} involves double infinite sum. However, we can 
consider the following expansion of the Green function which involves one infinite sum :  
\begin{equation}
\label{eq:ansatzG2}
\mathcal{G}_{q/p}(x,y|x_0,y_0) = \frac{4}{(p+q)^2} \sum_{l=1}^\infty 
\sum_{i,j=1}^2\left[ c_{0,l}^{i,j} a_0^{(i)} f_0^{(i)} (x) g_0^{(i)} (x_0) a_l^{(j)} f_l^{(j)} (y) g_l^{(j)} (y_0)\right] 
+ \frac{2}{p+q} \sum_{k=1}^\infty \sum_{i=1}^2 \left[a_k^{(1)} f_k^{(i)} (x) g_k^{(i)} (x_0) F^{(i)}_{2 \pi k}(y|y_0)\right],
\end{equation}
where the the coefficients $a_l^{(j)}$ and $c_{0,l}^{i,j}$  are given in \eqref{a_k} and \eqref{coeff-GF} respectively. 
The functions $F^{(1)}_{2 \pi k}(y|y_0)$ for $k\geq1$ are unknown functions to be determined. Putting this expansion of Green function in \eqref{eq:Gf} we find that 
the functions have to satisfy
\begin{eqnarray}
\label{eq:odeF}
F^{(1)}_{2 \pi k}(y|y_0)'' - (2 \pi k)^2 F^{(1)}_{2 \pi k}(y|y_0) &=& \delta(y-y_0), \\
F^{(2)}_{2 \pi k}(y|y_0)'' - (2 \pi k)^2 F^{(2)}_{2 \pi k}(y|y_0) &=& \delta(y-y_0) 
+ 8 (p-q) \pi k \frac{a_k^{(1)} g_k^{(1)}(x_0)}{a_k^{(2)} g_k^{(2)}(x_0)} F^{(1)}_{2 \pi k}(y|y_0). \nonumber
\end{eqnarray}
Solving Eqs.\eqref{eq:odeF} and simplifying gives 
\begin{eqnarray}
\label{eq:solF}
F^{(1)}_{2 \pi k}(y|y_0) &=& F_{2 \pi k}(y|y_0), \\
F^{(2)}_{2 \pi k}(y|y_0) &=& F_{2 \pi k}(y|y_0) + 8 (p-q) \pi k \frac{a_k^{(1)} g_k^{(1)}(x_0)}{a_k^{(2)} g_k^{(2)}(x_0)} \Sigma_{2 \pi k}(y|y_0), \nonumber
\end{eqnarray}
where $F_a(y|y_0)$ and $\Sigma_a(y|y_0)$ are given in \eqref{F_a} and \eqref{Sigma} respectively. 
For $q=0$, putting these results in the expansion \eqref{eq:ansatzG2} gives the solution~\eqref{Green-func}.

From \eqref{eq:ansatzG} one can also obtain \eqref{eq:ansatzG2} by using the following identity for $k>0$
\begin{eqnarray}
F_{2 \pi k}(y|y_0)=\sum_{m=0}^\infty \left( \frac{m \sin (2 \pi  m y) \cos (2 \pi  m y_0)}{\pi ^3 \left(k^2+m^2\right)^2}
-\frac{(1-y_0) \sin (2 \pi  m y) \sin (2 \pi  m y_0)+ a_m^{(1)}y \cos (2 \pi  m y) \cos (2 \pi  m y_0)}{\pi ^2
   \left(k^2+m^2\right)} \right). \nonumber
\end{eqnarray}


\begin{thebibliography}{10}

\bibitem{spohn1991}
H.~Spohn, {\em Large Scale Dynamics of Interacting Particles}.
\newblock Springer-Verlag, 1991.

\bibitem{bertini2015}
L.~Bertini, A.~D. Sole, D.~Gabrielli, G.~Jona-Lasinio, and C.~Landim,
  ``Macroscopic fluctuation theory,'' {\em Rev. Mod. Phys.}, vol.~87,
  pp.~593--636, 2015.

\bibitem{derrida2007}
B.~Derrida, ``Non-equilibrium steady states: fluctuations and large deviations
  of the density and of the current,'' {\em J. Stat. Mech.}, p.~P07023, 2007.

\bibitem{chou_m_z2011}
T.~Chou, K.~Mallick, and R.~K.~P. Zia, ``Non-equilibrium statistical mechanics:
  from a paradigmatic model to biological transport,'' {\em Rep. Prog. Phys..},
  vol.~74, p.~116601, 2011.

\bibitem{lazarescu2015}
A.~Lazarescu, ``The physicist's companion to current fluctuations:
  One-dimensional bulk-driven lattice gases,'' {\em arXiv:1507.04179v1}, 2015.

\bibitem{krug_g2000}
J.~Krug and J.~Garc{\'i}a, ``Asymmetric particle systems on $\mathbb{R}$,''
  {\em J. Stat. Phys.}, vol.~99, pp.~31--55, 2000.

\bibitem{schutz2000}
G.~M. Sch{\"u}tz, ``Exact tracer diffusion coefficient in the asymmetric random
  average process,'' {\em J.~Stat.~Phys.}, vol.~99, pp.~1045--1049, 2000.

\bibitem{rajesh_m2001}
R.~Rajesh and S.~N. Majumdar, ``Exact tagged particle correlations in the
  random average process,'' {\em Phys. Rev. E}, vol.~64, no.~036103, 2001.

\bibitem{feng_i_s1996}
S.~Feng, I.~Iscoe, and T.~Sepp{\"a}l{\"a}inen, ``A class of stochastic
  evolutions that scale to the porous medium equation,'' {\em J. Stat. Phys.},
  vol.~85, pp.~513--517, 1996.

\bibitem{coppersmith1996}
S.~N. Coppersmith, C.~h.~Liu, S.~Majumdar, O.~Narayan, and T.~A. Witten,
  ``Model for force fluctuations in bead packs,'' {\em Phys. Rev. E}, vol.~53,
  pp.~4673--4685, 1996.

\bibitem{rajesh_m2000}
R.~Rajesh and S.~N. Majumdar, ``Conserved mass models and particle systems in
  one dimension,'' {\em J. Stat. Phys.}, vol.~99, no.~3/4, pp.~943--965, 2000.

\bibitem{melzak1976}
Z.~A. Melzak, {\em Mathematical Ideas, Modeling and Applications, Vol II of
  Companion to Concrete Mathematics}.
\newblock Wiley, New York, 1976.

\bibitem{ferrari_f1998}
P.~A. Ferrari and L.~R.~G. Fontes, ``Fluctuations of a surface submitted to a
  random average process,'' {\em El. J. Prob.}, vol.~3, pp.~1--34, 1998.

\bibitem{ispolatov1998}
S.~Ispolatov, P.~L. Krapivsky, and S.~Redner, ``Wealth distributions in asset
  exchange models,'' {\em Eur. Phys. J. B}, vol.~2, pp.~267--276, 1998.

\bibitem{aldous_d1995}
D.~Aldous and P.~Diaconis, ``Hammersley's interacting particle process and
  longest increasing subsequences,'' {\em Probab. Theory Relat. Fields},
  vol.~103, pp.~199--213, 1995.

\bibitem{zielen_s2002a}
F.~Zielen and A.~Schadschneider, ``Exact mean-field solutions of the asymmetric
  random average process,'' {\em J. Stat. Phys.}, vol.~106, pp.~173--185, 2002.

\bibitem{zielen_s2003}
F.~Zielen and A.~Schadschneider, ``Matrix product approach for the asymmetric
  random average process,'' {\em J. Phys. A: Math. Gen.}, vol.~36,
  pp.~3709--3723, 2003.

\bibitem{zielen_s2002b}
F.~Zielen and A.~Schadschneider, ``Broken ergodicity in a stochastic model with
  condensation,'' {\em Phys. Rev. Lett.}, vol.~89, p.~090601, 2002.

\bibitem{gutsche2008}
C.~Gutsche, F.~Kremer, M.~Kr{\"u}ger, M.~Rauscher, R.~Weeber, and J.~Harting,
  ``Colloids dragged through a polymer solution: Experiment, theory, and
  simulation,'' {\em J. Chem. Phys.}, vol.~129, p.~084902, 2008.

\bibitem{kruger_r2009}
M.~Kr{\"u}ger and M.~Rauscher, ``Diffusion of a sphere in a dilute solution of
  polymer coils,'' {\em J. Chem. Phys.}, vol.~131, p.~094902, 2009.

\bibitem{candelier_d2010}
R.~Candelier and O.~Dauchot, ``Journey of an intruder through the fluidization
  and jamming transitions of a dense granular media,'' {\em Phys. Rev. E},
  vol.~81, p.~011304, 2010.

\bibitem{pesic2012}
J.~Pesic, J.~Z. Terdik, X.~Xu, Y.~Tian, A.~Lopez, S.~A. Rice, A.~R. Dinner, and
  N.~F. Scherer, ``Structural responses of quasi-two-dimensional colloidal
  fluids to excitations elicited by nonequilibrium perturbations,'' {\em Phys.
  Rev. E}, vol.~86, p.~031403, 2012.

\bibitem{dullens_b2011}
R.~P.~A. Dullens and C.~Bechinger, ``Shear thinning and local melting of
  colloidal crystals,'' {\em Phys. Rev. Lett.}, vol.~107, p.~138301, 2011.

\bibitem{burlatsky1992}
S.~F. Burlatsky, G.~S. Oshanin, A.~V. Mogutov, and M.~Moreau, ``Directed walk
  in a one-dimensional lattice gas,'' {\em Phys. Lett. A}, vol.~166,
  pp.~230--234, 1992.

\bibitem{burlatsky1996}
S.~F. Burlatsky, G.~Oshanin, M.~Moreau, and W.~P. Reinhardt, ``Motion of a
  driven tracer particle in a one-dimensional symmetric lattice gas,'' {\em
  Phys. Rev. E}, vol.~54, no.~4, pp.~3165--3172, 1996.

\bibitem{landim_o_v1998}
C.~Landim, S.~Olla, and S.~B. Volchan, ``Driven tracer particle in one
  dimensional symmetric simple exclusion,'' {\em Commun. Math. Phys.},
  vol.~192, pp.~287--307, 1998.

\bibitem{benichou1999}
O.~B{\'e}nichou, A.~M. Cazabat, A.~Lemarchand, M.~Moreau, and G.~Oshanin,
  ``Biased diffusion in a one-dimensional adsorbed monolayer,'' {\em J. Stat.
  Phys.}, vol.~97, pp.~351--371, 1999.

\bibitem{Beni_et_al2013}
O.~Benichou, P.~Illien, C.~Mejía-Monasterio, and G.~Oshanin, ``A biased
  intruder in a dense quiescent medium: looking beyond the force–velocity
  relation,'' {\em J. Stat. Mech : Theory and experiment}, p.~P05008, 2013.

\bibitem{evans_h2005}
M.~R. Evans and T.~Hanney, ``Nonequilibrium statistical mechanics of the
  zero-range process and related models,'' {\em J. Phys. A: Math. Gen.},
  vol.~38, pp.~R195--R239, 2005.

\bibitem{Evans_M_Z2004}
M.~R. Evans, S.~N. Majumdar, and R.~K.~P. Zia, ``Factorized steady states in
  mass transport models,'' {\em J. Phys. A: Math. Gen.}, vol.~37, p.~L275,
  2004.

\bibitem{zia_e_m2004}
R.~K.~P. Zia, M.~R. Evans, and S.~N. Majumdar, ``Construction of the factorized
  steady state distribution in models of mass transport,'' {\em J. Stat. Mech.
  : Theory and experiment}, p.~L10001, 2004.

\bibitem{Cha_P_M2015}
A.~Chatterjee, P.~Pradhan, and P.~K. Mohanty, ``Cluster-factorized steady
  states in finite range processes,'' {\em Phys. Rev. E}, vol.~92, p.~032103,
  2015.

\bibitem{majumdar2010}
``Real-space condensation in stochastic mass transport models,'' in {\em Les
  Houches lecture notes for the summer school on « Exact Methods in
  Low-dimensional Statistical Physics and Quantum Computing »}, Oxford
  university press, 2010.

\bibitem{rajesh_m_01}
R.~Rajesh and S.~N. Majumdar, ``Exact phase diagram of a model with aggregation
  and chipping,'' {\em Phys. Rev. E}, vol.~63, no.~036114, 2001.

\bibitem{Sha_m_m2014}
T.~Sadhu, S.~N. Majumdar, and D.~Mukamel, ``Long-range correlations in a
  locally driven exclusion process,'' {\em J. Stat. Phys}, vol.~90,
  pp.~012109--1, 2014.

\bibitem{bodineau_d_l2010}
T.~Bodineau, B.~Derrida, and J.~L. Lebowitz, ``A diffusive system driven by a
  battery or by a smoothly varying field,'' {\em J. Stat. Phys}, vol.~140,
  p.~648, 2010.

\end{thebibliography}

\end{document}